\documentclass{article}
\usepackage{graphicx} 

\usepackage{hyperref}
\usepackage{amsmath}
\usepackage{amssymb}
\usepackage{arydshln}
\usepackage{xcolor}
\usepackage{setspace}
\usepackage[top=20mm, bottom=20mm, left=25mm, right=25mm]{geometry}
\usepackage{color,soul} 
\usepackage{caption}
\usepackage{verbatim}
\newtheorem{theorem}{Theorem}

\title{Analysis of Stepped-Wedge Cluster Randomized Trials when treatment effects vary by exposure time or calendar time}
\author{Kenneth M. Lee,$^{1*}$ Elizabeth L. Turner,$^{2,3}$ Avi Kenny$^{2,3}$
\date{}
}

\begin{document}

\maketitle

\noindent $^1$Department of Biostatistics, Epidemiology and Informatics, University of Pennsylvania, Philadelphia, PA, USA

\noindent $^2$Department of Biostatistics \& Bioinformatics, Duke University, Durham, NC, USA

\noindent $^3$Duke Global Health Institute, Duke University, Durham, NC, USA
    
\noindent * Corresponding Author. Center for Clinical Epidemiology and Biostatistics, University of Pennsylvania School of Medicine, Blockley Hall, 423 Guardian Drive, Philadelphia, PA 19104

\noindent E-mail: kenneth.lee@pennmedicine.upenn.edu

\newpage

\begin{abstract}
Stepped-wedge cluster randomized trials (SW-CRTs) are traditionally analyzed with models that assume an immediate and sustained treatment effect. 
Previous work has shown that making such an assumption in the analysis of SW-CRTs when the true underlying treatment effect varies by exposure time can produce severely misleading estimates.
Alternatively, the true underlying treatment effect might vary by calendar time. Comparatively less work has examined treatment effect structure misspecification in this setting.
Here, we evaluate the behavior of the mixed effects model-based immediate treatment effect, exposure time-averaged treatment effect, and calendar time-averaged treatment effect estimators in different scenarios where they are misspecified for the true underlying treatment effect structure.
We show that the immediate treatment effect estimator is relatively robust to bias when estimating a true underlying calendar time-averaged treatment effect estimand.
However, when there is a true underlying calendar (exposure) time-varying treatment effect, misspecifying an analysis with an exposure (calendar) time-averaged treatment effect estimator can yield severely misleading estimates and even converge to a value of the opposite sign of the true calendar (exposure) time-averaged treatment effect estimand.
\textcolor{black}{In this article, we highlight the two different time scales on which treatment effects can vary in SW-CRTs and clarify potential vulnerabilities that may arise when considering different types of time-varying treatment effects in a SW design. Accordingly, we emphasize the need for researchers to carefully consider whether the treatment effect may vary as a function of exposure time and/or calendar time in the analysis of SW-CRTs.}
\end{abstract}

{\raggedright \textbf{Keywords:} stepped-wedge cluster randomized trial, time-varying treatment effects, exposure time, calendar time, model misspecification, estimands \par}
\break

\section{Introduction}

In cluster randomized trials (CRTs), clusters of individuals are randomized to treatment conditions (typically assigned to receive either the treatment or control). CRTs are often used when the treatment needs to be administered at the cluster level or when there is a risk of contamination between treatment conditions. The stepped-wedge cluster randomized trial (SW-CRT) is a specific type of CRT where all clusters begin the trial receiving the control and are randomized into sequences to start receiving the treatment at different time periods \cite{hussey_design_2007}. The staggered crossover is uni-directional and continues until all clusters are exposed to the treatment. This phased implementation may potentially be logistically advantageous, making the SW-CRT an increasingly popular design.

SW-CRTs have historically been analyzed with statistical models that implicitly assume the effect of the treatment is immediate and sustained over the entire trial duration \cite{hussey_design_2007},
but in reality, the treatment effect may vary with time. It is crucial to be precise when discussing such time-varying treatment effects, since there are two distinct time scales present in SW-CRTs, typically referred to as \textit{calendar time} and \textit{exposure time}. Calendar time is the amount of time that has passed since some fixed point in time. Calendar time is sometimes also referred to as \textit{study time}, where it refers to the amount of time passed since the start of the study. In contrast, exposure time, sometimes also referred to as \textit{duration time}, refers to the amount of time that has passed since the start of the initial treatment administration for a particular cluster. In a traditional parallel trial design in which all clusters begin treatment simultaneously, these time scales are equivalent \cite{lee_cluster_2024}. However, in a stepped wedge design, these time scales are distinct due to the staggered rollout of the treatment.

Treatment effects may vary by exposure time in scenarios where the treatment has a ``cumulative'' or a ``learning'' effect dependent on the length of time a cluster has has been receiving the treatment \cite{kenny_analysis_2022,lee_cluster_2024,maleyeff_assessing_2023}. Such exposure time-varying treatment effects have also been previously referred to as \textit{time-on-treatment} effects \cite{hughes_current_2015}.
For example, in a SW-CRT where clusters are hospital wards implementing a novel treatment, there may be a learning effect where physicians in each ward become more adept at administering the treatment over time since they first started administering it, leading to a more effective treatment with greater exposure time.
Replacement and supplementation SW-CRT designs can also be considered as examples of having such exposure time-varying treatment effects by design \cite{lyons_proposed_2017}.
Crucially, it has been shown that when exposure time-varying treatment effects are present in an SW-CRT, misspecification of the treatment effect structure with an immediate treatment effect can potentially yield incredibly misleading estimates that can even converge to the opposite sign of the true average treatment effect over exposure time \cite{kenny_analysis_2022}.

Treatment effects may also vary by calendar time \cite{hemming_analysis_2017}. This can take many different forms. Treatment effects may vary seasonally; for example, a treatment designed to reduce the incidence of heat stroke may have stronger effects during hotter months of the year. There may also be exogenous shocks to the entire study area that happen at a single point in time (i.e., an earthquake or disease outbreak) which alter the implementation (and therefore the effect) of a treatment. There can also be changes to the study population (or to the nature of the control condition) that affect the magnitude of the treatment effect, such as the standard of care rapidly improving over time. It may also be appropriate to interpret calendar time as study time in some adaptive or pragmatic trial designs where there may be changes to the nature of the treatment in later study periods resulting from insights gathered from earlier periods.

When it comes to variables that vary with calendar time, it is critical to distinguish between those that affect the outcome (separate from the treatment effect) and those that modify the treatment effect (i.e., a calendar time-varying treatment effect). The former are typically accounted for in SW-CRT analysis models by modeling calendar time with indicator variables corresponding to calendar time periods. With the latter, comparatively less methodological work has explored the properties of such underlying calendar time-varying treatment effects. More specifically, little methodological work has explored different models that account for such calendar-time varying treatment effects and the impact of model misspecification with such a treatment effect structure or when such an underlying treatment effect structure is present.

Previous work has demonstrated that treatment effect estimators in SW-CRTs can be robust to certain kinds of arbitrary model misspecification, including misspecification of the correlation structure, as long as the treatment effect structure is correctly specified \cite{wang_how_2024}. 
However, it is not always obvious whether a SW-CRT may have underlying exposure, calendar, or both (also referred to as \textit{saturated}) time-varying treatment effects.

In this current work, we describe how misspecification of the treatment effect structure in the analysis of SW-CRTs can yield biased estimators for the true time-averaged treatment effect estimands. We then quantify the resulting bias across different SW-CRT scenarios. We will focus on model misspecification when there is a true underlying exposure or calendar time-varying treatment effect, but not both. Therefore, scenarios with a true underlying saturated time-varying treatment effect are excluded and considered outside the scope of this current work.
In Section \ref{sect:estimands}, we first define different time-averaged treatment effect estimands of interest. In Section \ref{sect:models}, we define the mixed effects model-based time-averaged treatment effect estimators for these estimands. In Section \ref{sect:results}, we summarize previous work on exposure time-varying treatment effects by Kenny et al. \cite{kenny_analysis_2022} before exploring the bias in different misspecified estimators for the true time-averaged treatment effect estimands. 
We follow this up with a simulation study to confirm our analytic results (Section \ref{sect:simulation}), reanalyze data from an illustrative case study (Section \ref{sect:case study}), before ending with some concluding remarks (Section \ref{sect:discussion}).
\textcolor{black}{Altogether, in this article, we highlight the two time scales on which treatment effects may vary in SW-CRTs and emphasize the need for researchers to carefully consider whether the treatment effect may vary as a function of either exposure time or calendar time in the analysis of SW-CRTs.}

\section{Time-varying treatment effect estimands}
\label{sect:estimands}

In this section, we use a model-based approach to broadly describe the underlying data-generating process of cross-sectional SW-CRTs and to define the time-averaged treatment effect estimands of interest. \textcolor{black}{More details regarding the data-generating process are included in the Appendix \ref{sect:appendix_DGP}.} 
In a SW-CRT with assumed fixed cluster-period cell sizes $K_{ij}=K \,\forall 
 \, ij$, we observe outcomes $Y_{ijk}$ for individual $k \in \{1,...,K\}$ in time period $j \in \{1,...,J\}$ of cluster $i \in \{1,...,I\}$, randomized to sequence $q \in \{1,...,Q\}$, where the treatment in sequence $q$ is introduced during period $j=q+1$ in a SW-CRT. Note that there are typically $Q=J-1$ sequences in a standard complete SW-CRT design, with $I/Q$ clusters equally randomized into each sequence $q$.

As mentioned in the introduction, most statistical models used to analyze data from stepped wedge designs assume an immediate treatment effect. In the current work, we assume that the treatment effect may vary as a function of exposure time $s_{ij}$ or calendar time period $j$ (but not both). \textcolor{black}{Accordingly, we separately define data-generating processes for cross-sectional SW-CRTs with these three treatment effect structures (i.e., immediate/no time-varying treatment effect, exposure time-varying treatment effects or calendar time-varying treatment effects) and an identity link function.}
We broadly specify the expected outcome of $Y_{ijk}$ conditional on \textcolor{black}{$\Gamma_j$ being the calendar time trend, $C_{ijk}$ being the cluster-specific, time-specific, and/or individual-specific heterogeneity term which captures the correlation structure of the data \cite{li_mixed-effects_2021}, and} $\dot{X}_{ij}$, $\ddot{X}_{ij}'$, or $\dddot{X}_{ij}'$ representing the immediate, exposure time-varying, or calendar time-varying treatment effect structures, respectively (equations \ref{eq:DGP3.1} - \ref{eq:DGP3.3}).

\textcolor{black}{Equation \ref{eq:DGP3.1} represents a general model with an immediate treatment effect structure. Where $q_i$ is the SW-CRT sequence assignment for cluster $i$, then $\dot{X}_{ij} = I(j>q_i)$ is the indicator ($=1$ if assigned to treatment, $=0$ if assigned to control) for the (single) immediate treatment effect $\theta$ for the individuals observed during period $j$ of cluster $i$, and}:
\begin{align}
\label{eq:DGP3.1}
    E[Y_{ijk}|\dot{X}_{ij},\Gamma_j,C_{ijk}] &= \dot{X}_{ij}\theta + \Gamma_j + C_{ijk} \,.
\end{align}
\textcolor{black}{Equation \ref{eq:DGP3.2} represents a general model with an exposure time-varying treatment effect structure. For simplicity of notation, we use the subscript $s\in(1,2,...,J-1)$ to represent the treatment exposure time, with $s_{ij}$ corresponding to that of cluster $i$ at time $j$. Accordingly, $\ddot{X}_{ij}'=\left(I(s_{ij}=1), I(s_{ij}=2),...,I(s_{ij}=J-1)\right)$ is the 1 by $J-1$ row vector of indicatorscorresponding to $\delta=(\delta_1,\delta_2,...,\delta_{J-1})'$ as the $J-1$ by 1 column vector of the different exposure time-varying treatment effects $\delta_s$ for individuals observed during period $j$ of cluster $i$, and}:
\begin{align}
\label{eq:DGP3.2}
    E[Y_{ijk}|\ddot{X}_{ij}',\Gamma_j,C_{ijk}] &= \ddot{X}_{ij}'\delta + \Gamma_j + C_{ijk} \,.
\end{align}
\textcolor{black}{Equation \ref{eq:DGP3.3} represents a general model with a calendar time-varying treatment effect structure. Accordingly, $\dddot{X}_{ij}' = \left(I(j=2 \ \& \ j>q_i),I(j=3 \ \& \ j>q_i),...,I(j=J \ \& \ j>q_i)\right)$ is the 1 by $J-1$ row vector of indicators corresponding to $\xi=(\xi_2,\xi_3,...,\xi_{J})'$ as the $J-1$ by 1 column vector of the different calendar time-varying treatment effects $\xi_j$ for individuals observed during period $j$ of cluster $i$, and}:
\begin{align}
\label{eq:DGP3.3}
    E[Y_{ijk}|\dddot{X}_{ij}',\Gamma_j,C_{ijk}] &= \dddot{X}_{ij}'\xi + \Gamma_j + C_{ijk} \,.
\end{align}
\textcolor{black}{Notably, equation \ref{eq:DGP3.1} is a special case of equations \ref{eq:DGP3.2} and \ref{eq:DGP3.3}, where $\delta_s=\theta \, \forall \, s$ and $\xi_j=\theta \, \forall \, j$, respectively. More details are included in the Appendix \ref{sect:appendix_DGP}.}

\begin{figure}[htp]
    \centering
    \includegraphics[width=12cm]{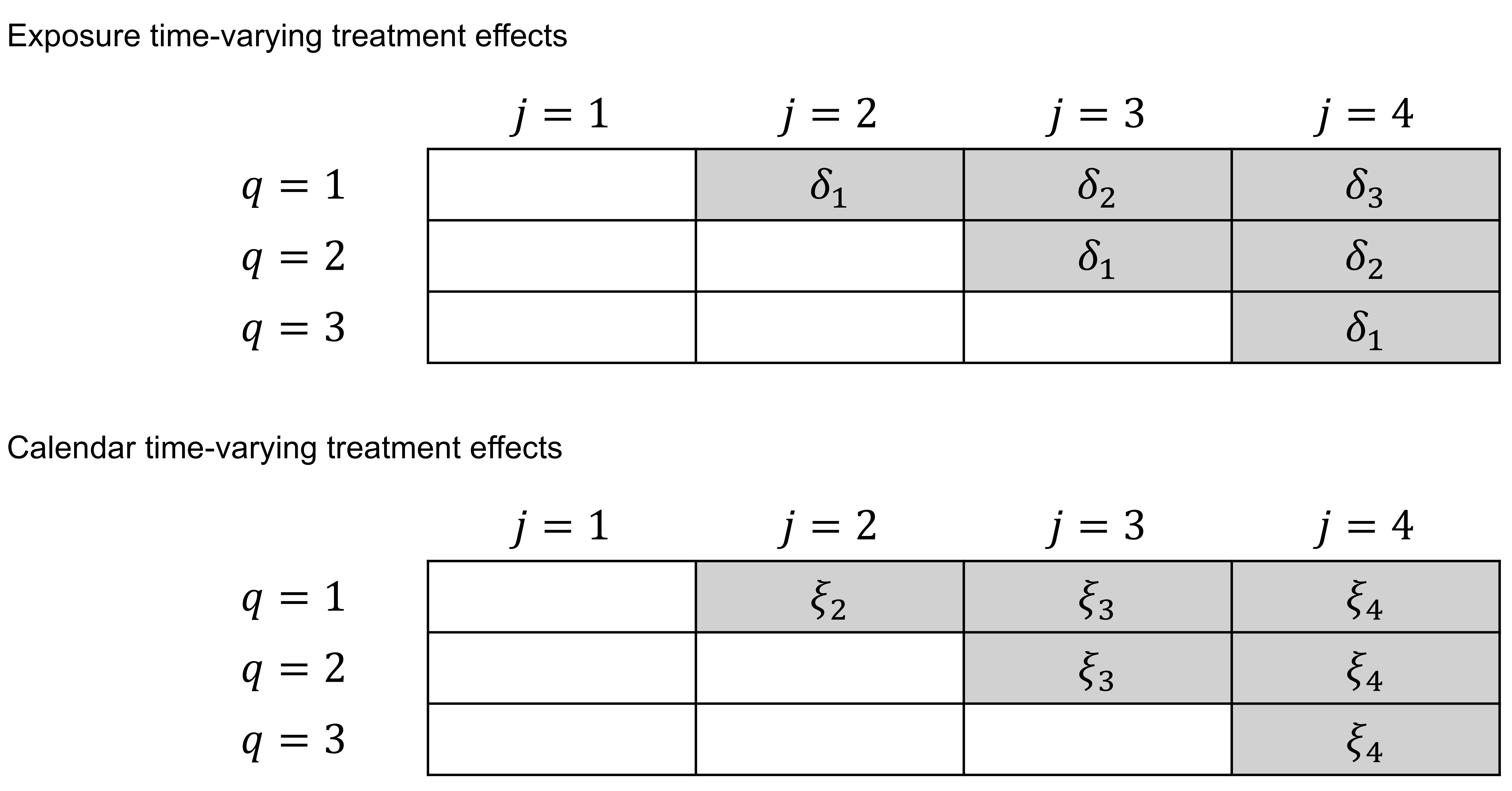}
    \caption{A 3 sequence, 4 period SW-CRT with exposure time-varying treatment effects $\delta_s$ and calendar time-varying treatment effects $\xi_j$.}
    \label{fig:Figure_Example_Design}
\end{figure}

\begin{figure}[htp]
    \centering
    \includegraphics[width=12cm]{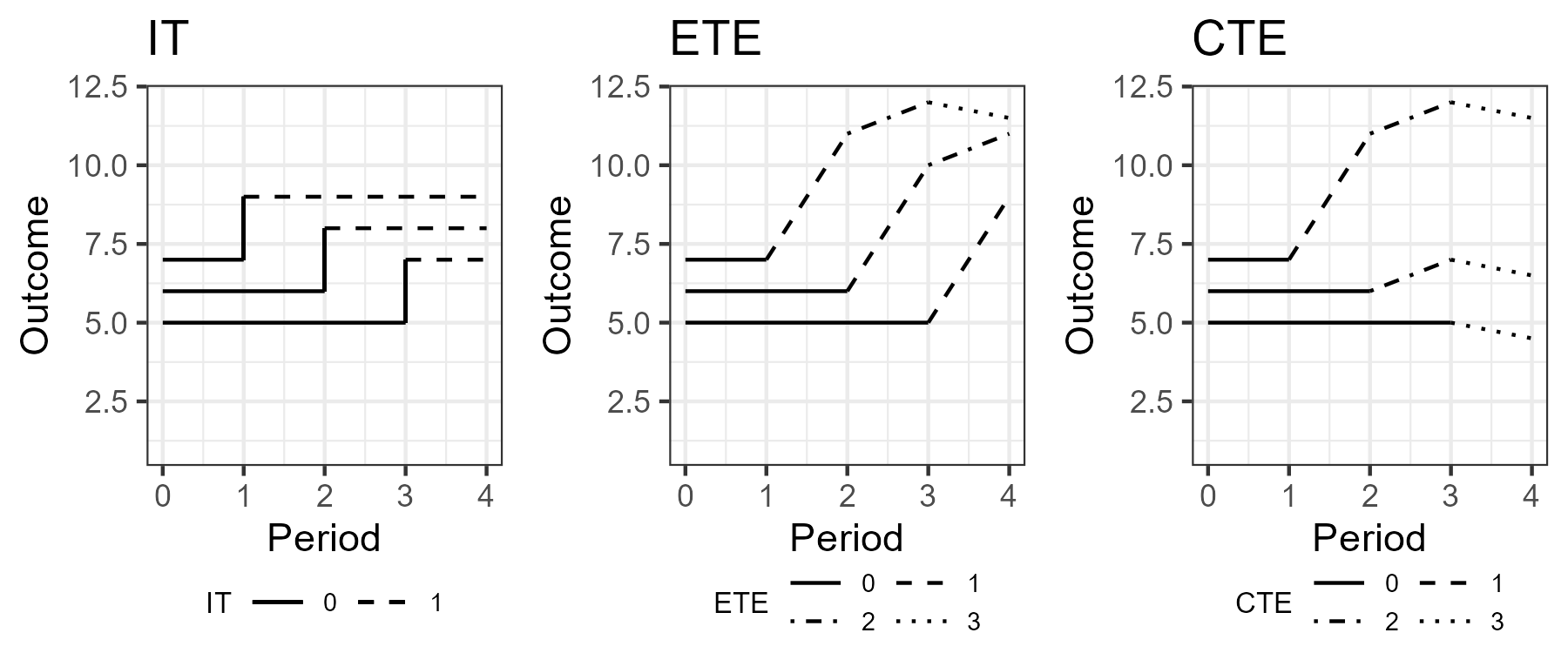}
    \caption{An example of immediate (IT), exposure time-varying (ETE), and calendar time-varying (CTE) treatment effects \textcolor{black}{plotted across calendar time (referred to on the x-axis as ``Period")} in a 4 period SW-CRT design. Period 0 corresponds with the start of the trial and period 1 corresponds with the end of period $j=1$. \textcolor{black}{Each line represents a different cluster, with solid line-segments representing periods when clusters are assigned to the control. Subsequent different line-types correspond with potential time-varying treatment effects corresponding to the treatment effect structure.}}
    \label{fig:Figure_Example_Effects}
\end{figure}

\textcolor{black}{For simplicity, we assume that only one of equations \ref{eq:DGP3.1} - \ref{eq:DGP3.3} are true.} Although in theory, the treatment effect may vary as a function of \textit{both} exposure and calendar time scales, we make the simplifying assumption in the current work that if treatment does vary over time, it only varies along one scale. The quantity $\delta_s$ can then be interpreted as the treatment effect at exposure time $s$, and the quantity $\xi_j$ can be interpreted as the treatment effect at calendar time $j$. We emphasize that all of these quantities are defined in terms of continuous time, even though time is often measured discretely in SW-CRTs. The difference between these time scales is illustrated in Figure \ref{fig:Figure_Example_Design}, with an example of such time-varying treatment effects in Figure \ref{fig:Figure_Example_Effects} for a simple 3 sequence, 4 period SW-CRT. 

The exposure time-varying treatment effect curve defined by the function $s \mapsto \delta_s$ or the calendar time-varying treatment effect curve defined by the function $j \mapsto \xi_j$ will often be of interest, as they completely characterize how the treatment effect varies with time, assuming the correct treatment effect structure is specified in a standard SW-CRT.

However, in the context of a randomized trial, it is typically desirable to have a single scalar estimand that can be used as the basis for testing the null hypothesis that the treatment is not effective on average over the duration of the trial. One such estimand previously considered in Kenny et al. \cite{kenny_analysis_2022} is the ``time-averaged treatment effect'', which we will refer to as the \textit{exposure time-averaged treatment effect} (ETATE). The ETATE between exposure times $s_1$ and $s_2$ can be defined as:
\[ETATE_{[s_1,s_2]} \equiv \frac{1}{s_2-s_1} \int_{s_1}^{s_2}\delta_s\textit{ds} \,.\]

\noindent The corresponding estimand for the model in which the treatment effect varies with calendar time is termed the \textit{calendar time-averaged treatment effect} (CTATE) \textcolor{black}{and has previously been described in Wang et al. \cite{wang_how_2024} and Chen \& Li \cite{chen_model-assisted_2025}}. The CTATE between calendar times $j_1$ and $j_2$ can be analogously defined as:
\[CTATE_{[j_1,j_2]} \equiv \frac{1}{j_2-j_1} \int_{j_1}^{j_2}\xi_j\textit{dj} \,.\]

\noindent \textcolor{black}{Both the ETATE and CTATE estimands are defined above as integrals, as in reality the exposure and calendar time scales are continuous. However, in many SW-CRT designs, discrete-time data are collected instead of continuous-time data, and some authors prefer to define analogous estimands as summations of the finite set of model parameters for simplicity (Section \ref{sect:models}).} Note that the ETATE is only well-defined if there are no calendar-time varying treatment effects $\xi_j=0 \, \forall \, j$, and the CTATE is only well-defined if there are no exposure-time varying treatment effects $\delta_s=0 \, \forall \, s$. To reiterate, if the immediate treatment effect assumption is correct, then by definition, there exists a scalar $\theta$ such that $\delta_s=\xi_j=\theta \, \forall \, s,j$. Only in this case does it make sense to define the \textit{immediate} treatment effect (IT) estimand equal to $\theta$.

In SW-CRTs with fixed time period lengths, the time-averaged treatment effect estimands over the course of the study will often be of interest, defined as $ETATE_{[0,J-1]}$ and $CTATE_{[1,J-1]}$ for the two time scales present in a SW-CRT. In this current work, we will primarily focus on these two estimands.
For convenience, we will refer to these estimands as the ETATE and CTATE, respectively, and drop the subscript denoting the interval.
For the ETATE estimand, the lower bound \textcolor{black}{of the integral} is the exposure time period $s_1=0$, corresponding with the start of exposure to the treatment, and the upper bound is $s_2=J-1$ in a standard SW-CRT design.
For the CTATE, the lower bound \textcolor{black}{of the integral} is calendar time period $j_1=1$, since the treatment is not introduced until the end of the first period (or equivalently, the start of the second period), and the effect of the treatment is not estimable at any calendar time prior to the introduction of the treatment, unless further assumptions are made (i.e., a parametric form for the function $j \mapsto \xi_j$). The upper bound for CTATE is chosen to be $j_2=J-1$, since after this time, all clusters will be in the treatment condition. As a result, without further assumptions, the discretely modelled calendar time effects and the calendar time-varying treatment effects are not separable $\forall \, j > J-1$ \cite{wang_how_2024}. We will further clarify this point in Section \ref{sect:models} and re-emphasize this point throughout this present article.

\section{Analytic models}
\label{sect:models}

In this section, we describe several mixed effects models that can be used to analyze data from a SW-CRT in which data is generated according to equations \ref{eq:DGP3.1} - \ref{eq:DGP3.3}. Some researchers prefer the generalized estimating equation (GEE) framework instead of the mixed effects model framework using generalized least squares (GLS). Notably, with an identity link function, the GLS estimator of such a linear mixed effects model has the same form as the GEE estimator \cite{gardiner_fixed_2009,hubbard_gee_2010}. In this current work, we focus on such linear analyses of continuous outcomes and maintain this equivalence.

We will assume discrete time periods, as is standard in the analysis of SW-CRTs and for simplicity. Accordingly, the calendar time trend is modeled using $P_j$ representing the 1 by $J$ row vector of indicators for each time period $j$, with the corresponding $J$ by 1 vector of period effects $\phi=(\phi_1,...,\phi_j,...,\phi_J)'$ (i.e., $\Gamma_j=P_j\phi$). The heterogeneity term is simply modeled as a cluster-level random intercept (i.e., $C_{qijk}=\alpha_i$ with $\alpha_1,...,\alpha_I \overset{iid}{\sim} N(0,\tau_{\alpha}^2)$). 

First we describe the immediate treatment effect (IT) model, given by:
\begin{equation}
\label{eq:ITmodel}
Y_{ijk}=\dot{X}_{ij}\theta+P_j\phi+\alpha_{i}+\epsilon_{ijk} \,,
\end{equation}

\noindent where the $\epsilon_{ijk} \overset{iid}{\sim} N(0,\sigma_{e}^2)$ terms are the model residuals, assumed to be mean-zero, identically distributed, and independent of each other and the $\alpha_i$ random effect terms. The IT model as described in equation \ref{eq:ITmodel} was introduced by Hussey and Hughes \cite{hussey_design_2007} and is appropriate to use if researchers are confident that the effect of the treatment does not vary with exposure or calendar time. 

Subsequently, the exposure time indicator (ETI) model is given by:
\begin{equation}
\label{eq:ETImodel}
Y_{ijk}=\ddot{X}_{ij}'\delta+P_j\phi+\alpha_{i}+\epsilon_{ijk} \,.
\end{equation}

\noindent This ETI model (equation \ref{eq:ETImodel}) has been previously described in Kenny et al. \cite{kenny_analysis_2022} and Lee and Cheung \cite{lee_cluster_2024}, and is appropriate if researchers think the treatment effect varies with exposure time. 

Finally, the calendar time indicator (CTI) model is given by:
\begin{equation}
\label{eq:CTImodel}
Y_{ijk}=\dddot{X}_{ij}'\xi+P_j\phi+\alpha_{i}+\epsilon_{ijk} \,.
\end{equation}

\noindent The CTI model (equation \ref{eq:CTImodel}) has been previously described by several authors, including Hemming et al. \cite{hemming_analysis_2017}, and is appropriate to use if researchers think the treatment effect varies with calendar time or study time. To reiterate, we use the convention that $\xi_J=0$, otherwise the CTI model as described in equation \ref{eq:CTImodel} is not identifiable (since the columns in the design matrix corresponding to $\xi_J$ and $\phi_J$ are identical); this was previously mentioned at the end of Section \ref{sect:estimands}.

The IT estimator is simply $\widehat{IT} = \widehat{\theta}$ from the IT model (equations \ref{eq:ITmodel}). With the ETI and CTI models (equation \ref{eq:ETImodel} \& \ref{eq:CTImodel}), the corresponding ETATE and CTATE estimators can be simply calculated as the unweighted means:

\[\widehat{ETATE}=\frac{\sum_{s=1}^{J-1}\hat{\delta}_s}{J-1}\]

\noindent and:

\[\widehat{CTATE}=\frac{\sum_{j=2}^{J-1}\hat{\xi}_j}{J-2}\]

\noindent in a standard SW-CRT design.

The generalized least squares (GLS) point estimator with an exchangeable correlation structure can be derived using the cluster-period means:
\[\hat{\beta}=(Z'V^{-1}Z)^{-1}ZV^{-1}Y\]

\noindent where $Z$ is the design matrix of all the fixed effects (including the treatment effect structure and period effect indicators), $V=\mathbb{I}_I \bigotimes R_i$ is the $IJ$ by $IJ$ block diagonal variance-covariance matrix of $Y$ (where $\mathbb{I}_I$ is an $I$ by $I$ dimension identity matrix), with $Y$ being the vector of cluster-period mean outcomes $\bar{Y}_{ij}$, and:
\[R_i = Var(\bar{Y}_{ij})(\mathbb{I}_J(1-\gamma) + \mathbb{J}_j(\gamma))\]
(where $\mathbb{I}_J$ and $\mathbb{J}_J$ are $J$ by $J$ dimension matrices, representing the identity matrix and a matrix of ones, respectively) with $\bigotimes$ representing the Kronecker product. Assuming fixed cluster-period sizes $K$, $Var(\bar{Y}_{ij})$ is equal to $\tau^{2}_{\alpha} + \sigma^2_e/K$ with an exchangeable correlation structure 
and cancels out of the generalized least squares (GLS) point estimator $(Z'V^{-1}Z)^{-1}ZV^{-1}Y$. 
The estimators can accordingly be written as a function of $\gamma$, where $\gamma=\frac{\tau^{2}_{\alpha}}{\tau^{2}_{\alpha} + \sigma^2_e/K}$. 

We acknowledge that the exchangeable correlation structure induced by the $\alpha_i$ random effects in equations \ref{eq:ITmodel} - \ref{eq:CTImodel} might not hold in many datasets, but we focus on these models for their simplicity.
In the Discussion (Section \ref{sect:discussion}), we describe how the results presented in this paper can be easily extended to models with a nested exchangeable correlation structure \cite{hooper_sample_2016,girling_statistical_2016}.

We will also consider the IT (equation \ref{eq:ITmodel}), ETI (equation \ref{eq:ETImodel}), and CTI models (equation \ref{eq:CTImodel}) with a specified independence correlation structure ($\tau^2_\alpha=0, \gamma=0, Var(\bar{Y}_{ij})=\sigma^2_e/K$), with coefficients accordingly determined using ordinary least squares (OLS).
This model was previously described in Matthews and Forbes \cite{matthews_stepped_2017} and resembles the SW-CRT vertical estimator \cite{thompson_robust_2018}.
Building on the equivalence between GEE and OLS with an identity link \cite{gardiner_fixed_2009,hubbard_gee_2010}, these marginal models are expected to yield unbiased estimates regardless of the specified correlation structure, assuming correct specification of the treatment effect structure \cite{ouyang_maintaining_2024,wang_how_2024}.

It is crucial to understand the (1.) distinction between data-generating models and their corresponding estimands versus (2.) analytic models and their corresponding estimators. For example, if there is a true underlying exposure time-varying treatment effect, the ETATE estimand will be well-defined but the CTATE estimand will not be. However, data resulting from such a data-generating model can still be analyzed using the IT (equation \ref{eq:ITmodel}), ETI (equation \ref{eq:ETImodel}), and CTI (equation \ref{eq:CTImodel}) models. An ETI model (equation \ref{eq:ETImodel}) can be used in this situation to derive an ETATE estimator, which (under certain conditions) is expected to be consistent and yield unbiased estimates for the true ETATE estimand. However, we can still erroneously analyze the data with a CTI model and derive the corresponding CTATE estimator to yield estimates of the CTATE estimand, even though the true CTATE estimand is not well-defined. Similarly, we might analyze the data using an immediate treatment (IT) model and produce an estimator of the IT estimand, which as discussed in Kenny et al. \cite{kenny_analysis_2022} is not well-defined if the immediate treatment assumption is incorrect.

\section{Behavior of models with misspecified time-varying treatment effect structures}
\label{sect:results}

In this section, we derive the bias in different time-averaged treatment effect estimators resulting from misspecified treatment effect structures. All results are derived assuming complete and balanced SW-CRT designs with equal allocation of clusters to each sequence $q$ and equal cluster-period cell sizes $K$.
\textcolor{black}{For simplicity, we assume that the true variance components $\tau^2_\alpha$ and $\sigma^2_e$ are known (as described in Hussey \& Hughes \cite{hussey_design_2007}), allowing the treatment effect estimators to be derived with generalized least squares \cite{wooldridge_econometric_2010}. This can be the case when conducting pre-trial power analyses \cite{hussey_design_2007}. Otherwise, feasible generalized least squares can be used with the estimated variance components, $\hat{\tau}^2_\alpha$ and $\hat{\sigma}^2_e$, when variance components are unknown \textit{a priori} \cite{wooldridge_econometric_2010}.}

\begin{figure}[htp]
    \centering
    \includegraphics[width=12cm]{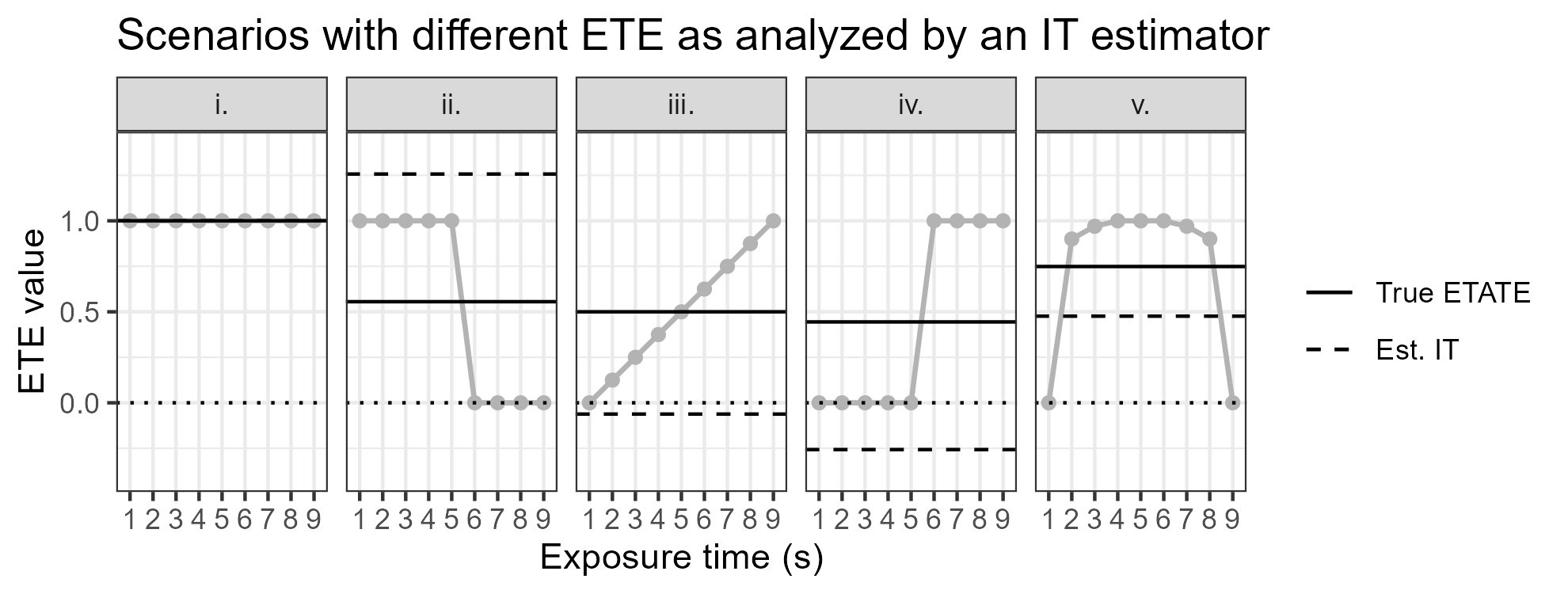}
    \caption{9 cluster, 10 period SW-CRT scenarios with different exposure time-varying treatment effect curves shown in gray. These different exposure time-varying treatment effects (ETE) are analyzed by an immediate treatment effect (IT) estimator. The correctly specified true exposure time-averaged treatment effect (True ETATE) and misspecified estimated immediate treatment effect (Est. IT) values are shown with the solid and dashed lines, respectively. The dotted black line marks an effect of 0.}
    \label{fig:Figure_Scenarios_IT_ETE}
\end{figure}

\begin{figure}[htp]
    \centering
    \includegraphics[width=12cm]{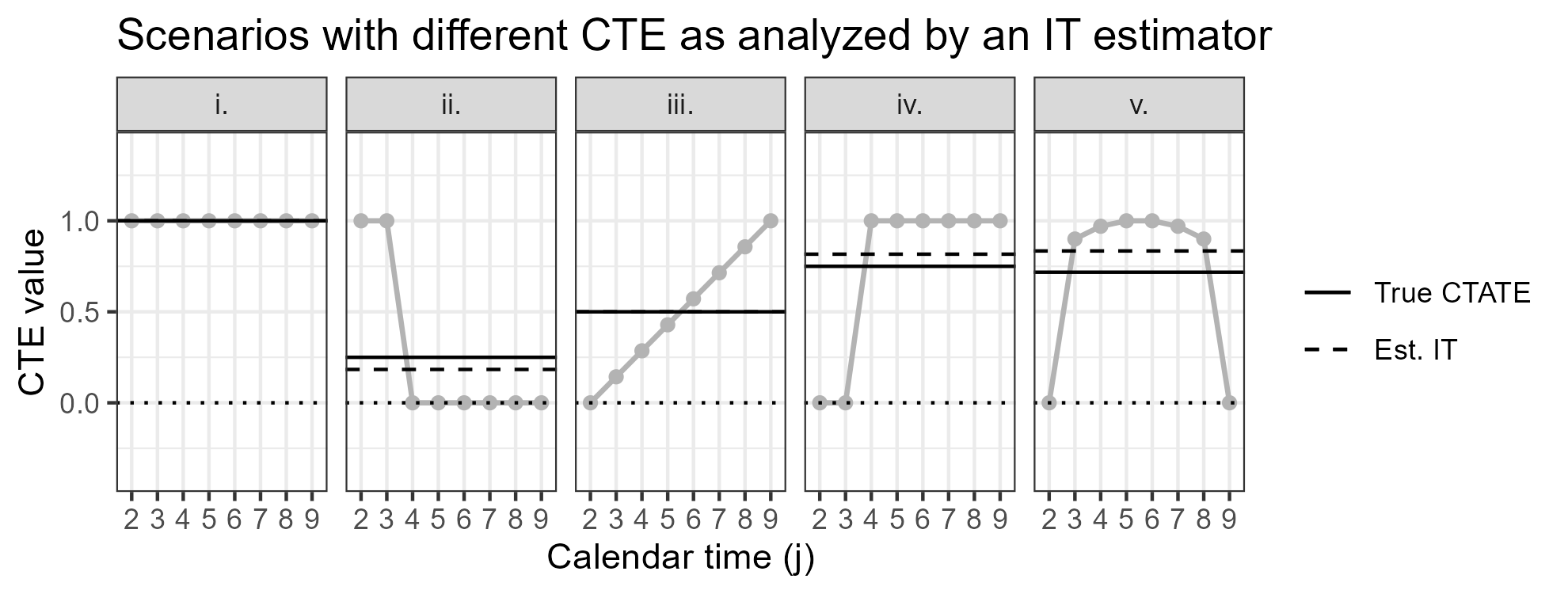}
    \caption{9 cluster, 10 period SW-CRT scenarios with different time-varying treatment effect curves shown in gray. These different calendar time-varying treatment effects (CTE) are analyzed by an immediate treatment effect. The correctly specified true calendar time-averaged treatment effect (True CTATE) and misspecified immediate treatment effect (Est IT) values are shown with the solid and dashed lines, respectively. The dotted black line marks an effect of 0. To reiterate, CTE are only identifiable up to time period 9 (excluding period 10) with the inclusion of period fixed effects.}
    \label{fig:Figure_Scenarios_IT_CTE}
\end{figure}

\begin{figure}[htp]
    \centering
    \includegraphics[width=12cm]{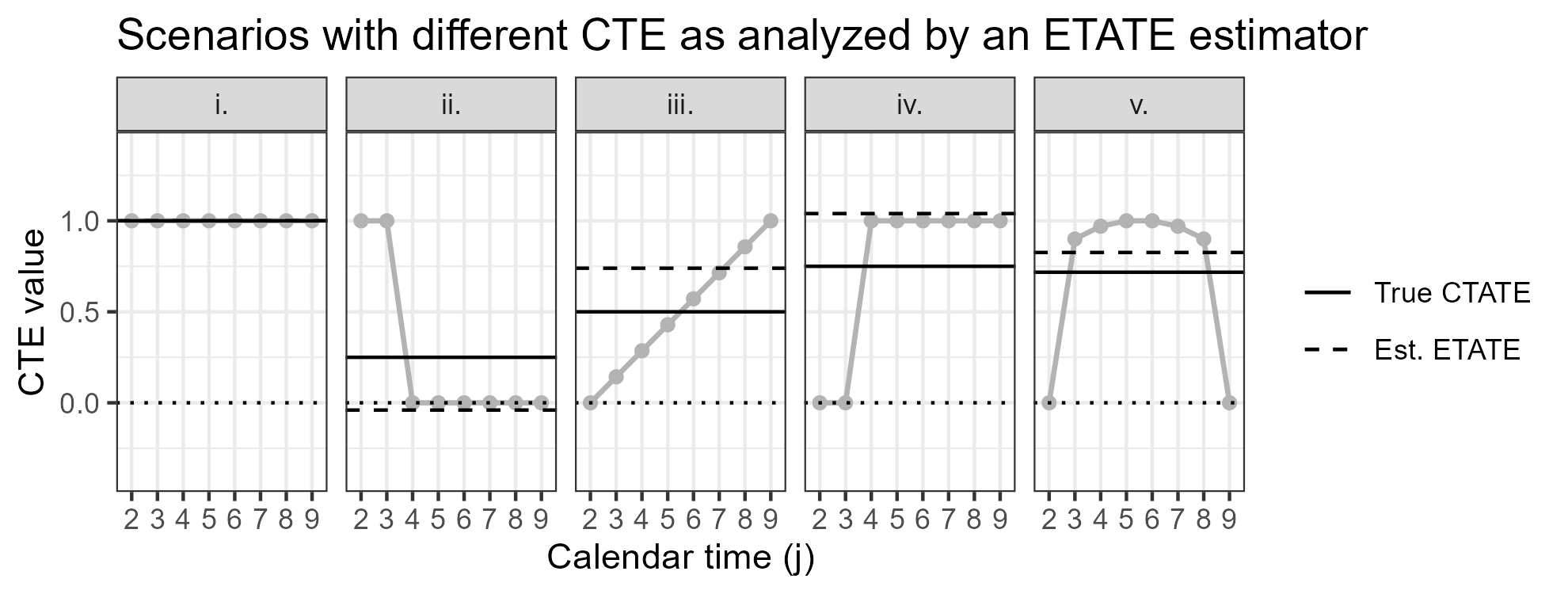}
    \caption{9 cluster, 10 period SW-CRT scenarios with different time-varying treatment effect curves shown in gray. These different CTE are analyzed by an exposure time-averaged treatment effect (ETATE). The correctly specified true calendar time-average treatment effect (True CTATE) and misspecified estimated exposure time-averaged treatment effect (Est. ETATE) values are shown with the solid and dashed lines, respectively. The dotted black line marks an effect of 0. To reiterate, CTE are only identifiable up to time period 9 (excluding period 10) with the inclusion of period fixed effects.}
    \label{fig:Figure_Scenarios_ETATE_CTE}
\end{figure}

\begin{figure}[htp]
    \centering
    \includegraphics[width=12cm]{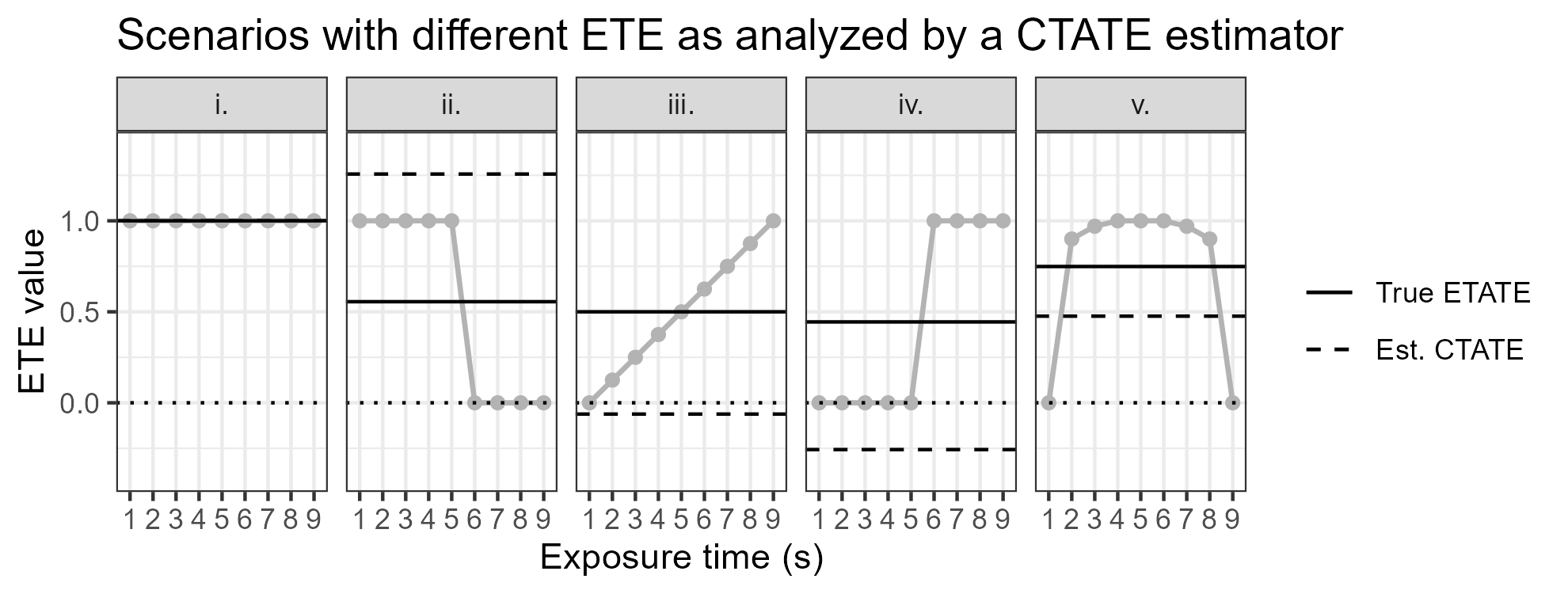}
    \caption{9 cluster, 10 period SW-CRT scenarios with different time-varying treatment effect curves shown in gray. These different ETE are analyzed by a calendar time-averaged treatment effect (CTATE). The correctly specified true exposure time-averaged treatment effect (True ETATE) and misspecified estimated calendar time-averaged treatment effect (Est. CTATE) values are shown with the solid and dashed lines, respectively. The dotted black line marks an effect of 0. To reiterate, CTE are only identifiable up to time period 9 (excluding period 10) with the inclusion of period fixed effects.}
    \label{fig:Figure_Scenarios_CTATE_ETE}
\end{figure}

To illustrate the influence of misspecifying the treatment effect structures, several possible exposure and calendar time-varying treatment effect curves are displayed in Figures \ref{fig:Figure_Scenarios_IT_ETE}, \ref{fig:Figure_Scenarios_IT_CTE}, \ref{fig:Figure_Scenarios_ETATE_CTE}, \& - \ref{fig:Figure_Scenarios_CTATE_ETE}, with the true and estimated time-averaged treatment effects shown with solid and dashed lines, respectively. The exposure and calendar time-varying treatment effect curves in Figures \ref{fig:Figure_Scenarios_IT_ETE}, \ref{fig:Figure_Scenarios_IT_CTE}, \ref{fig:Figure_Scenarios_ETATE_CTE}, \& \ref{fig:Figure_Scenarios_CTATE_ETE} are included to illustrate specific scenarios where misspecified treatment effect estimators are expected to yield more misleading results. We discuss these figures in detail in relevant sections below.

\subsection{Behavior of the IT estimator with a true underlying exposure time-varying treatment effect structure}
\label{sect:IT_ETATE}

First, we summarize some of the results from Kenny et al. \cite{kenny_analysis_2022}, demonstrating that the immediate treatment effect (IT) estimator as described in equation \ref{eq:ITmodel} is a weighted sum of the exposure time-varying treatment effect estimands, with some weights potentially being negative.

As previously shown in Kenny et al. \cite{kenny_analysis_2022}, the IT estimator can be written as:

\begin{equation}
\label{eq:IT}
\widehat{IT}= 
\frac{12(1+\gamma Q)}{Q(Q+1)(\gamma Q^2+2Q-\gamma Q-2)}\sum_{j=1}^{J}\sum_{q=1}^{Q}\left[Q(I(j>q))-j+1+\frac{\gamma Q(2q-Q-1)}{2(1+\gamma Q)}\right]\Bar{Y}_{qj}
\end{equation}

\noindent with average outcomes $\Bar{Y}_{qj}$ from sequence $q\in \{1,...,Q\}$ during calendar time $j \in \{1,...,J\}$, where $\gamma=\frac{\tau^{2}_{\alpha}}{\tau^{2}_{\alpha} + \sigma^2_e/K}$ \cite{kenny_analysis_2022}. $I(j>q)$ is an indicator for whether the index for a given period $j$ is higher than that of a given sequence $q$.

Assuming the true underlying marginal model has exposure time-varying treatment effects, as shown below:
\[E[\Bar{Y}_{ij}|\ddot{X}_{ij}',P_j]=\ddot{X}_{ij}'\delta + P_j\phi \,.\]
Kenny et al. \cite{kenny_analysis_2022} use equation \ref{eq:IT} to demonstrate that the expected value of the immediate treatment effect estimator is:

\begin{equation}
\label{eq:E[IT]_ETE}
E[\widehat{IT}|\ddot{X}_{ij}',P_j]=
\sum_{s=1}^{J-1} w_{1}(Q,\gamma,s) \delta_s \,,
\end{equation}

\noindent with estimand weights corresponding to specific exposure time-varying treatment effect estimands:

\begin{equation}
w_{1}(Q,\gamma,s) \equiv
\frac{6(s-Q-1)((1+2\gamma Q)s-(1+\gamma+\gamma Q)Q)}{Q(Q+1)(\gamma Q^2+2Q-\gamma Q-2)} \,.
\end{equation}

\noindent Based on the above result, the conditional expectation of the IT estimator can potentially converge to a negative number despite $\delta_s \geq 0 \, \forall \, s$.

In Figure \ref{fig:Figure_Scenarios_IT_ETE}, we display several possible exposure time-varying treatment effect curves along with the true ETATE estimand value and the conditionally expected IT estimate. As previously demonstrated in Kenny et al. \cite{kenny_analysis_2022}, the potentially negative weights $w_{1}(Q,\gamma,s)$ can yield estimates that are severely underestimated or overestimated for the true ETATE estimand, with some estimates being negative despite the true exposure time-varying treatment effect estimands all being $\delta_s \geq 0 \, \forall \, s$ (Figure \ref{fig:Figure_Scenarios_IT_ETE}).

Notably, when $\gamma=0$, which corresponds to an IT estimator with an independence correlation structure:
\[
w_{1}(Q,s)=
\left(\frac{6(s-Q-1)(s-Q)}{Q(Q+1)(2Q-2)}\right) \,.
\]
We observe that the above $w_{1}(Q,s)\geq 0 \, \forall \, Q, s$ in standard SW-CRT designs where $s \in \{1,...,J-1\}$ and $Q=J-1$. Accordingly, the IT estimator with an independence correlation structure will often be less biased for the ETATE estimand compared to corresponding analyses with an exchangeable correlation structure where $\gamma > 0$.

\subsection{Behavior of the IT estimator with a true underlying calendar time-varying treatment effect structure}
\label{sect:IT_CTATE}

Building on the work by Kenny et al. \cite{kenny_analysis_2022}, we demonstrate that the immediate treatment effect (IT) estimator can be written as a weighted sum of the calendar-time varying treatment effect estimands when the true model has calendar time varying treatment effects:

\begin{theorem}
\label{theorem:IT_CTE}
Assuming a standard SW-CRT where the true underlying marginal model has calendar time-varying treatment effects, as shown below:
\[E[\Bar{Y}_{ij}|\dddot{X}_{ij}',P_j]=\dddot{X}_{ij}'\xi+P_j\phi \,,\]
the misspecified immediate treatment (IT) estimator, as described in equation \ref{eq:ITmodel}, is in conditional expectation a weighted average of the true calendar time-varying treatment effect estimands $\xi_j$:

\begin{equation}
\label{eq:E[IT]_CTE}
E[\widehat{IT}|\dddot{X}_{ij}',P_j]=
\sum_{j=2}^{J-1}w_2(Q,j)\xi_j \,,
\end{equation}

\noindent where $w_2(Q,j)$ are weights corresponding with calendar time period $j$:

\begin{equation}
w_2(Q,j) \equiv \frac{6(j-1)(Q+1-j)}{Q(Q+1)(Q-1)} \,.
\end{equation}

\end{theorem}

\noindent A proof of Theorem \ref{theorem:IT_CTE} is provided in Appendix \ref{sect:appendix_derivation_IT_CTE}.

Notably, the weights for the calendar time-varying treatment effects $\xi_j$ in analyses with an exchangeable correlation structure are independent of $\gamma$ and never negative $w_2(Q,j) \geq 0 \, \forall \, Q,j$. Accordingly, the IT estimator can yield less biased results when there is an underlying calendar time-varying treatment effect structure, in contrast to the results presented in Section \ref{sect:IT_ETATE} for scenarios with a true underlying exposure time-varying treatment effect structure (where the IT estimator is shown to be a weighted sum of exposure time-varying treatment effect estimands with $\gamma$-dependent weights that can potentially be negative).

In the appendix (Appendix \ref{sect:appendix_weights}), we plot the scaled weights $(J-2)[w_{2}(Q,j)]$ in the IT estimator for each calendar time-varying treatment effect estimand $\xi_j$, across different size SW-CRTs.
Somewhat surprisingly, the immediate treatment effect is an unbiased estimator of the CTATE estimand in a 3 sequence, $J=4$ period SW-CRT (Appendix \ref{sect:appendix_weights}).

In Figure \ref{fig:Figure_Scenarios_IT_CTE}, we display several different scenarios with a true underlying calendar time-varying treatment effect, along with the true CTATE estimand value and the conditionally expected IT estimate. 
Overall, we observe that misspecifying the analysis of calendar time-varying treatment effects with an immediate treatment effect can still yield estimates that are fairly close to the original estimate (Figure \ref{fig:Figure_Scenarios_IT_CTE}).

In specific scenarios where some calendar time-varying treatment effect estimands are positive and some negative over the duration of the trial, the immediate treatment effect estimator can still yield an estimate with the opposite sign of the true CTATE estimand. We include an example of this in Appendix \ref{sect:appendix_additional_scenarios_IT_CTE}. While these scenarios are unlikely to be common in practice, researchers should still be aware of this possibility.

\subsection{Behavior of the ETATE estimator with a true underlying calendar time-varying treatment effect structure}
\label{sect:ETATE_CTATE}

In this section, we explore the results of using a (misspecified) ETI model (equation \ref{eq:ETImodel}) for analysis in scenarios with true underlying calendar time-varying treatment effects. \textcolor{black}{More information is provided in Appendix \ref{sect:appendix_derivation_ETATE_CTE}.}

\begin{figure}[htp]
    \centering
    \includegraphics[width=12cm]{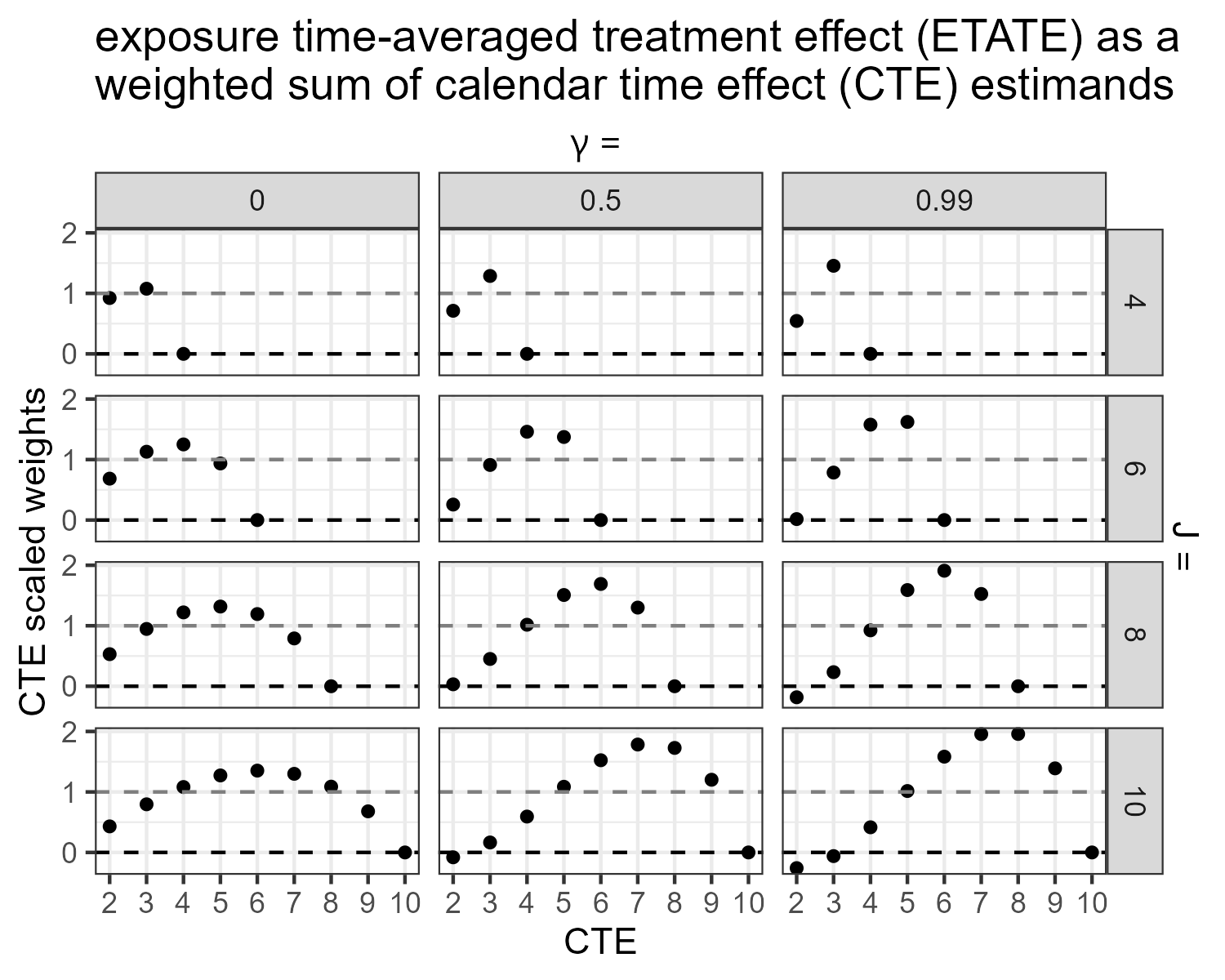}
    \caption{Calendar time-varying treatment effects and their corresponding scaled weights $(J-2)[w_3(Q, \gamma, j)]$ in the ETATE estimator are graphed on the x-axis and y-axis, respectively. Results are presented across SW-CRTs with different total numbers of periods $J$ and varying $\gamma=\frac{\tau^{2}_{\alpha}}{\tau^{2}_{\alpha} + \sigma^2_e/K}$. The dashed gray line marks a nominal weight of 1. The dashed black line marks a weight of 0.}
    \label{fig:ETATE_CTE}
\end{figure}

In a $Q=J-1$ sequence, $J$ period SW-CRT, with $I/Q$ clusters per sequence, where the true underlying marginal model has calendar time-varying treatment effects (CTI treatment effect structure) as shown below:
\[E[\Bar{Y}_{ij}|\dddot{X}_{ij}',P_j]=\dddot{X}_{ij}'\xi+P_j\phi \,,\]
the conditional expectation of the exposure time indicator (ETI) estimators $\hat{\delta}_s$ described in equation \ref{eq:ETImodel}, given the above marginal model with $\dddot{X}_{ij}'$ for the true CTI treatment effect structure, is then:
\[
E[\hat{\delta}_s|\dddot{X}_{ij}',P_j]=\sum_{i=1}^{I}\sum_{j=1}^{J}\lambda_{[\delta_s]ij}E[\bar{Y}_{ij}|\dddot{X}_{ij}',P_j]
\]
with $\lambda_{[\delta_s]ij}$ corresponding to weights for outcomes observed in period $j$ of cluster $i$ for each exposure time-varying treatment effect $\delta_s$. These weights are difficult to express in closed-form but can be easily computed numerically. Still, we can demonstrate that this simplifies to:
\[
E[\hat{\delta}_s|\dddot{X}_{ij}',P_j]=\sum_{j=2}^{J-1}\sum_{q=1}^{j-1}\lambda_{[\delta_s]qj}\xi_j \,.
\]
Therefore, individual ETI estimators $\hat{\delta}_s$ are in expectation, weighted sums of the CTI estimands.
Subsequently, the misspecified exposure time-averaged treatment effect (ETATE) is in conditional expectation a weighted average of the true calendar time-varying treatment effect estimands $\xi_j$:
\begin{align*}
    E[\widehat{ETATE}|\dddot{X}_{ij}',P_j] = \frac{\sum_{s}^{J-1}E[\hat{\delta}_s|\dddot{X}_{ij}',P_j]}{J-1} = \sum_{j=2}^{J-1}w_3(Q,\gamma,j)\xi_j 
\end{align*}

\noindent where estimand weights $w_3(Q,\gamma,j)=\frac{1}{J-1}\left(\sum_{s=1}^{J-1}\sum_{q=1}^{j-1}\lambda_{[\delta_s]qj}\right)$ vary by the total number of sequences $Q=J-1$ and $\gamma=\frac{\tau^{2}_{\alpha}}{\tau^{2}_{\alpha} + \sigma^2_e/K}$ with equal cluster-period size $K$ and an exchangeable correlation structure, as demonstrated numerically (Figure \ref{fig:ETATE_CTE}).

Specifically, in a 3 sequence, 4 period SW-CRT, we can demonstrate that:
\[
\begin{split}E[\widehat{ETATE}|\dddot{X}_{ij}',P_j] &= \frac{E[\hat{\delta}_1+\hat{\delta}_2+\hat{\delta}_3|\dddot{X}_{ij}',P_j]}{3} = w_3(Q,\gamma,2)\xi_2 + w_3(Q,\gamma,3)\xi_3 \\
&=
\frac{-9\gamma^2+30\gamma+12}{2(9\gamma^2+39\gamma+13)}\xi_2+\frac{27\gamma^2+48\gamma+14}{2(9\gamma^2+39\gamma+13)}\xi_3 \,.
\end{split}
\]
 
In Figure \ref{fig:ETATE_CTE}, we plot the scaled weights $(J-2)[w_3(Q,\gamma,j)]$ in the ETATE estimator for each exposure time-varying treatment effect $\xi_j$, across different size SW-CRTs and values of $\gamma$ as derived with numerical matrix inversion (with more scenarios included in Appendix \ref{sect:appendix_weights}).

Notably in Figure \ref{fig:ETATE_CTE}, larger SW-CRT designs with higher intracluster correlation (ICC) values and large cluster-period sample sizes $K$ (and therefore higher values of $\gamma$) can yield increasingly extreme scaled weights. In larger SW-CRT designs, the weights for earlier calendar time-varying treatment effect estimands $\xi_j$ can be negative with larger values of $\gamma$.

In Figure \ref{fig:Figure_Scenarios_ETATE_CTE}, we display several different scenarios with a true underlying calendar time-varying treatment effect, along with the true CTATE estimand value and the conditionally expected ETATE estimate. With the potentially negative weights, the ETATE estimate can be severely underestimated or overestimated, with some estimates being negative despite the true calendar time-varying treatment effect estimands all being $\xi_j \geq 0$ (Figure \ref{fig:Figure_Scenarios_ETATE_CTE}).

In contrast, in the explored conditions where $\gamma=0$, which is the case when specifying an independence correlation structure, the weights in the ETATE estimator are all $w_3(Q,\gamma,j) \geq 0 \, \forall \, Q, \gamma, j$ and can yield less biased results, despite the treatment effect structure misspecification (Figure \ref{fig:ETATE_CTE}).

\subsection{Behavior of the CTATE estimator with a true underlying exposure time-varying treatment effect structure}
\label{sect:CTATE_ETATE}

In this section, we explore the results of using a (misspecified) CTI model for analysis in scenarios where there are underlying exposure time-varying treatment effects. \textcolor{black}{More information is provided in Appendix \ref{sect:appendix_derivation_CTATE_ETE}.}

\begin{figure}[htp]
    \centering
    \includegraphics[width=12cm]{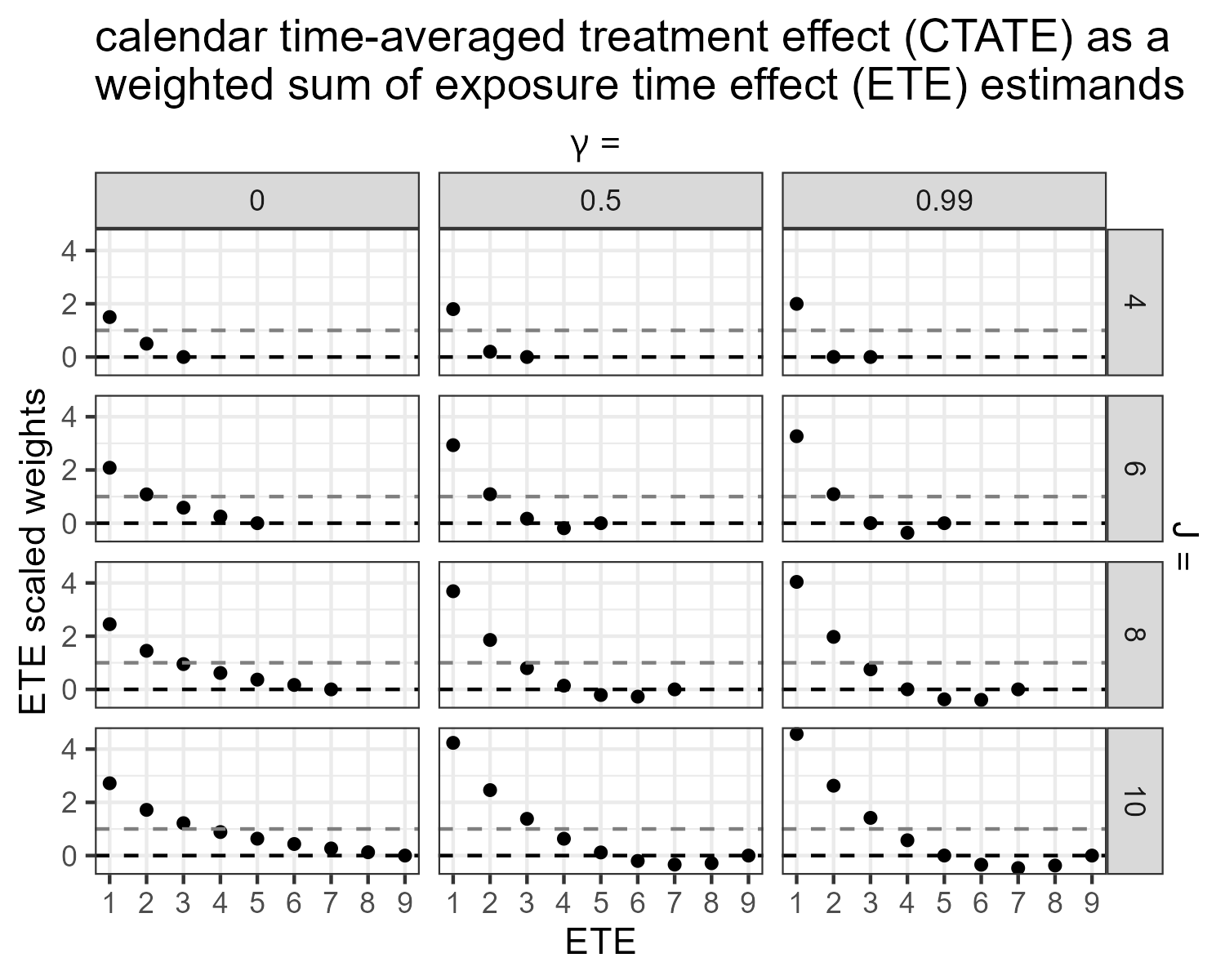}
    \caption{Exposure time-varying treatment effects and their corresponding scaled weights $(J-2)w_4(Q,\gamma,s)$ in the CTATE estimator are graphed on the x-axis and y-axis, respectively. Results are presented across SW-CRTs with different total numbers of periods $J$ and varying $\gamma=\frac{\tau^{2}_{\alpha}}{\tau^{2}_{\alpha} + \sigma^2_e/K}$. The dashed gray line marks a nominal weight of 1. The dashed black line marks a weight of 0.}
    \label{fig:CTATE_ETE}
\end{figure}

In a $Q=J-1$ sequence, $J$ period SW-CRT, with $I/Q$ clusters per sequence, where the true underlying marginal model has exposure time-varying treatment effects (ETI treatment effect structure) as shown below:
\[
E[\Bar{Y}_{ij}|\ddot{X}_{ij}',P_j]=\ddot{X}_{ij}'\delta + P_j\phi \,,
\]
the conditional expectation of the calendar time indicator (CTI) estimators $\hat{\xi}_{j=c} \, \forall \, c \in \{2,...,J-1\}$ described in equation \ref{eq:CTImodel}, given the above marginal model with $\ddot{X}_{ij}'$ for the true ETI treatment effect structure, is then: \[
E[\hat{\xi}_{j=c}|\ddot{X}_{ij}',P_j]=\sum_{i=1}^{I}\sum_{j=1}^{J-1}\lambda_{[\xi_{j=c}]ij} E[\Bar{Y}_{ij}|\ddot{X}_{ij}',P_j]
\] 
with $\lambda_{[\xi_{j=c}]ij}$ corresponding to weights for outcomes observed in period $j$ of cluster $i$ for each calendar time-varying treatment effect $\xi_{j=c}$. These weights are difficult to express in closed-form but can be computed numerically. Still, we can demonstrate that this simplifies to:
\[
E[\hat{\xi}_{j=c}|\ddot{X}_{ij}',P_j]=\sum_{j=2}^{J-1}\sum_{q=1}^{j-1} \lambda_{[\xi_{j=c}]qj}\delta_{j-q} \,.
\]
Therefore, individual CTI estimators $\hat{\xi}_{j=c} \, \forall \, c \in \{2,...,J-1\}$ are in expectation each weighted sums of the ETI estimands.
Subsequently, the misspecified calendar time-averaged treatment effect (CTATE), as described in equation \ref{eq:CTImodel}, excluding information from period $J$, due to $\xi_J$ being unidentifiable with period effect $\phi_J$, is in conditional expectation a weighted average of the true calendar time-varying treatment effect estimands $\delta_s$:
\begin{align*}
    E[\widehat{CTATE}|\ddot{X}_{ij}',P_j] = \frac{\sum_{c=2}^{J-1}E[\hat{\xi}_{j=c}|\ddot{X}_{ij}',P_j]}{J-2} = \sum_{s=1}^{J-2} w_{4}(Q,\gamma,s) \delta_s
\end{align*}
where estimand weights $w_{4}(Q,\gamma,s) =\frac{1}{J-2}\left(\sum_{c=2}^{J-1}\sum_{j=2}^{J-1} \sum_{q=1}^{J-1} I(j-q=s)\lambda_{[\xi_j]qj}\right)$ (with $I(j-q=s)$ referring to an indicator for indices where the difference between the given period $j$ and sequence $q$ is equal to the given exposure time $s$, $j-q=s$) vary by total number of sequences $Q=J-1$, and $\gamma=\frac{\tau^{2}_{\alpha}}{\tau^{2}_{\alpha} + \sigma^2_e/K}$ (with equal cluster-period size $K$ and an exchangeable correlation structure) as demonstrated numerically (Figure \ref{fig:CTATE_ETE}).

Specifically, in a 3 sequence, 3 period SW-CRT (where we exclude period 4 due to the identifiability issues mentioned earlier) we can demonstrate that:
\[
\begin{split}E[\widehat{CTATE}|\ddot{X}_{ij}',P_j] &=\frac{E[\hat{\xi}_2+\hat{\xi}_3|\ddot{X}_{ij}',P_j]}{2} = w_{4}(Q,\gamma,1) \delta_1 + w_{4}(Q,\gamma,2) \delta_2 \\
&=\frac{9\gamma^2+15\gamma+6}{2(3\gamma^2+8\gamma+4)}\delta_1+\frac{-3\gamma^2+\gamma+2}{2(3\gamma^2+8\gamma+4)}\delta_2 \,.
\end{split}
\]

In Figure \ref{fig:CTATE_ETE}, we plot the scaled weights $(J-2)[w_4(Q,\gamma,s)]$ in the CTATE estimator for each exposure time-varying treatment effect estimand $\delta_s$, across different size SW-CRTs and values of $\gamma$ ($\gamma=\frac{\tau^{2}_{\alpha}}{\tau^{2}_{\alpha} + \sigma^2_e/K}$) as derived with numerical matrix inversion (with more scenarios included in Appendix \ref{sect:appendix_weights}).

Similar to Figure \ref{fig:ETATE_CTE}, Figure \ref{fig:CTATE_ETE} reveals that larger SW-CRT designs and higher ICC values can yield increasingly extreme scaled weights. Notably, in larger SW-CRT designs, the weights for the later exposure time-varying treatment effect estimands $\delta_s$ can be negative with larger values of $\gamma$. 

In Figure \ref{fig:Figure_Scenarios_CTATE_ETE}, we display some different scenarios with a true underlying exposure time-varying treatment effect, along with the true ETATE estimand value and the conditionally expected CTATE estimate. With such misspecifcation of the treatment effect structure, the CTATE estimate can be severely underestimated or overestimated, with some estimates being negative despite the true exposure time-varying treatment effect estimands all being $\delta_s \geq 0 \, \forall \, s$ (Figure \ref{fig:Figure_Scenarios_CTATE_ETE}).

Again, in the explored conditions where $\gamma=0$, which is the case when specifying an independence correlation structure, the ETE estimand weights in the CTATE estimator are all $w_{4}(Q,\gamma,s) \geq 0 \, \forall \, Q, \gamma, s$, which can reduce some bias despite the treatment effect structure misspecification (Figure \ref{fig:CTATE_ETE}).

\section{Simulation}
\label{sect:simulation}

We conducted a simulation study to confirm the analytic results reported in Section \ref{sect:results}. Data for balanced and complete SW-CRTs with equal allocation to each sequence were simulated with data generating processes from the following three scenarios:
\begin{enumerate}
\item $Y_{ijk}=\dot{X}_{ij}\theta+P_j\phi+\alpha_{i}+\epsilon_{ijk}$
\item $Y_{ijk}=\ddot{X}_{ij}'\delta+P_j\phi+\alpha_{i}+\epsilon_{ijk}$
\item $Y_{ijk}=\dddot{X}_{ij}'\xi+P_j\phi+\alpha_{i}+\epsilon_{ijk}$
\end{enumerate}
with (1.) a scalar immediate treatment effect estimand $\theta$, (2.) a $J-1$ by 1 column vector of exposure time-varying treatment effect estimands $\delta=(\delta_1,...,\delta_s,...,\delta_{J-1})'$, or (3.) a $J-2$ by 1 column vector of calendar time-varying treatment effect estimands $\xi=(\xi_2,...,\xi_j,...,\xi_{J-1})'$.

Simulations involved $I=18$ clusters, $J=10$ periods, and a constant $K=30$ individuals per cluster-period cell. We set cluster and residual variances, $\tau^2_\alpha=0.\Bar{1}$ and $\sigma^2_e=1$, where $\alpha_i \overset{iid}{\sim} N(0,\tau^2_\alpha)$ and $\epsilon_{ijk} \overset{iid}{\sim} N(0,\sigma^2_e)$, to generate an ICC of $\rho=\frac{\tau^{2}_{\alpha}}{\tau^{2}_{\alpha} + \sigma^2_e} =0.1$ and $\gamma=\frac{\tau^{2}_{\alpha}}{\tau^{2}_{\alpha} + \sigma^2_e/K} \approx 0.77$. We generated a linear calendar time trend, $\phi=(\phi_1,...,\phi_j,...,\phi_{10})=(5,6,7,8,9,10,11,12,13,14)'$ for periods $j\in[1,10]$.

With (1.) an immediate treatment effect structure, we set $IT=\theta=6$. With (2.) an exposure time-varying treatment effect structure, we set $\delta=(\delta_1,...,\delta_s,...,\delta_9)'=(0,0,0.5,1,2,4,6,6,6)'$ for exposure times $s\in[1,9]$, yielding a true ETATE estimand of $2.8\Bar{3}$. With (3.) a calendar time-varying treatment effect structure, we set $\xi=(\xi_2,...,\xi_j,...,\xi_9)'=(6,3,1,0.5,0.1,0,0,0)'$ for calendar times $j\in[2,9]$, yielding a true CTATE estimand of $1.325$. Recall that the calendar time-varying treatment effect in the final period $\xi_J=\xi_{10}$ is unidentifiable due to perfect collinearity with final period effect $\phi_J=\phi_{10}$.

All simulated data sets were then analyzed with the IT model (equation \ref{eq:ITmodel}), ETI model (equation \ref{eq:ETImodel}), or CTI model (equation \ref{eq:CTImodel}), with either an exchangeable or independence correlation structure, to produce the IT, ETATE, and CTATE estimators. 
The exposure and calendar time-varying treatment effect curves are graphed alongside the resulting IT, ETATE, and CTATE estimates from the models with different correlation structures (Appendix \ref{sect:appendix_simulation}).

Model-based variance estimators may be underestimated when the incorrect correlation structure is specified. Accordingly, inference is also performed using cluster robust variance estimators \cite{ouyang_maintaining_2024}. We will compare inference using the model-based variance estimator, the ``approximate jackknife'', or the bias-reduced linearization robust variance estimators from the ``clubSandwich'' R package, following previously published recommendations by Ouyang et al. \cite{ouyang_maintaining_2024}.

In each simulation scenario, we generated $1000$ simulated datasets. We present the results in terms of percent relative bias (\%), precision as the reciprocal of the average estimated variances, and coverage probability (CP) as the probability that the 95\% confidence interval contains the true estimand.
Precision and coverage probability in analyses with an independence correlation structure are additionally calculated with the model-based variance estimator, ``approximate jackknife'', and the bias-reduced linearization cluster robust variance estimators. We include the Monte Carlo SE's as the standard deviation of the 1000 simulated estimates in Appendix \ref{sect:appendix_simulation}. All simulations were conducted using R version 4.3.2 and structured using the SimEngine simulation package \cite{kenny_simengine_2024}, with simulation code available at \url{https://github.com/Avi-Kenny/Code__SW-CalTime}.

\subsection{Simulation results}
\begin{figure}[htp]
    \centering
    \includegraphics[width=12cm]{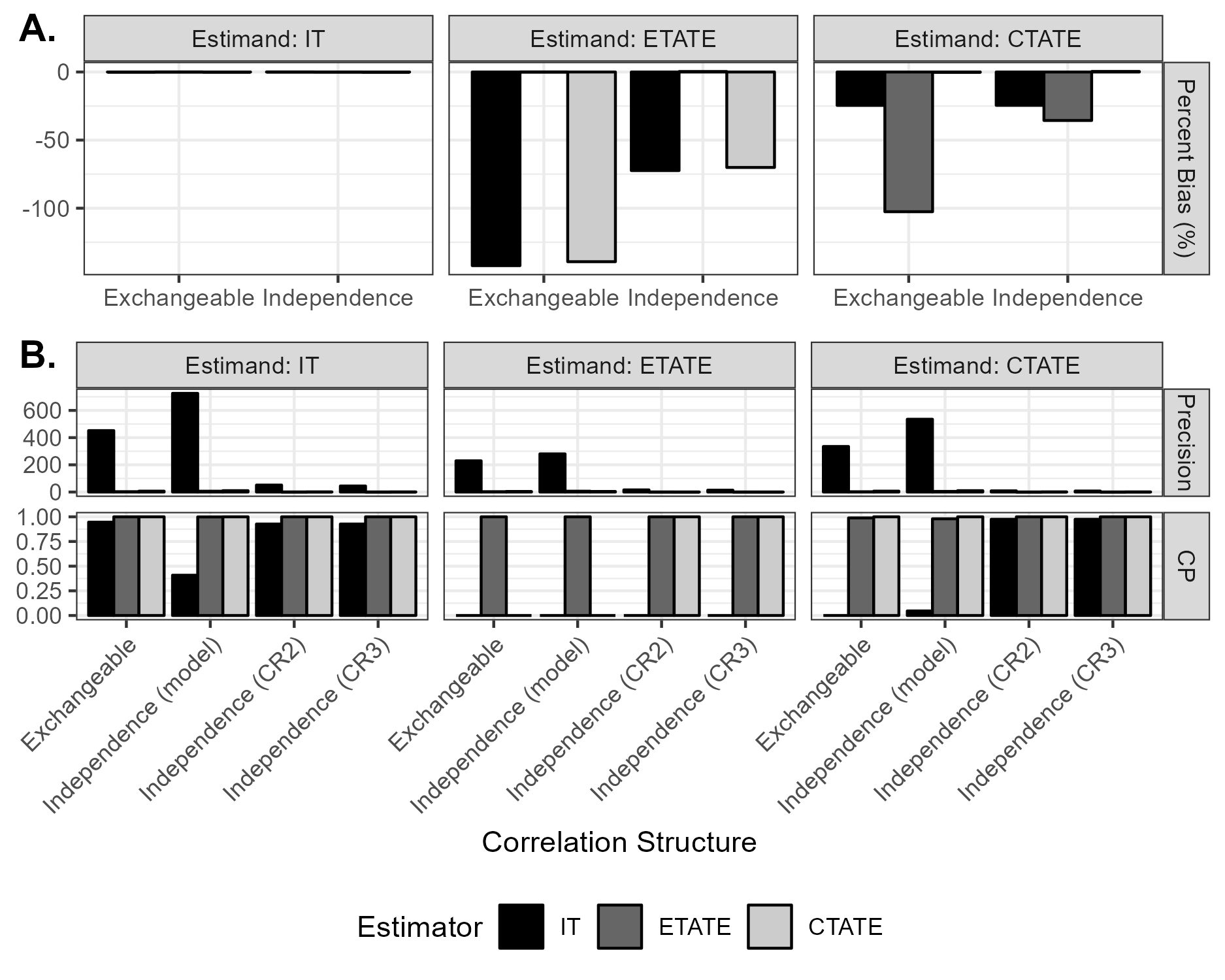}
    \caption{Simulation results in terms of percent bias, precision, and coverage probability (CP) for immediate treatment effect (IT), exposure time-averaged treatment effect (ETATE), and calendar time-averaged treatment effect (CTATE) estimands. Corresponding results from the $\widehat{IT}$, $\widehat{ETATE}$, and $\widehat{CTATE}$ estimators from analyses with an exchangeable or independence correlation structure are plotted. Precision and CP in analyses with an independence correlation structure are calculated with the model-based (model) variance estimator, bias-reduced linearization (CR2) cluster robust variance estimator, and ``approximate jackknife'' (CR3) cluster robust variance estimator. Analysis results are summarized over 1000 simulated replicates.}
    \label{fig:sim_results}
\end{figure}

As expected, all estimators produced unbiased estimates in simulation scenario 1 with a true underlying immediate treatment effect (Figure \ref{fig:sim_results}.A).

Analyses with misspecified treatment effect structures and an exchangeable correlation structure typically yielded biased results.
With a true underlying exposure time-varying treatment effect structure (simulation scenario 2), IT and CTATE estimators yielded severely biased results for the ETATE estimand (Figure \ref{fig:sim_results}.A), with the mean estimates being negative despite all the true exposure time-varying treatment effects being $\delta_s \geq 0 \, \forall \, s$. Similarly, with a true underlying calendar time-varying treatment effect structure (simulation scenario 3), the ETATE estimator yielded severely biased results for the CTATE estimand (Figure \ref{fig:sim_results}.A), with the mean estimates being negative despite all the calendar time-varying treatment effects being $\xi_j \geq 0 \, \forall \, j$.

In Section \ref{sect:results}, we previously observed in the explored conditions where $\gamma=0$ that weights of the ETATE and CTATE estimator are all $\geq 0$, which can produce less biased results, despite the treatment effect structure misspecification. Accordingly, we observe that the models with misspecified treatment effect structures and an independence correlation structure generally yielded considerably less biased estimates than the corresponding mixed effects model across all simulation scenarios (Figure \ref{fig:sim_results}.A).

In analyses with an exchangeable correlation structure, the IT estimator is much more precise than the ETATE and CTATE estimators. However, due to this precision, this estimator can have incredibly poor coverage probability when misspecified for the true underlying time-varying treatment effect structure (Figure \ref{fig:sim_results}.B). 
Surprisingly, despite the IT estimator being less biased than the ETATE estimator in scenarios with true underlying calendar time-varying treatment effects (Figure \ref{fig:sim_results}.A), its high precision resulted in it having practically no coverage of the 95\% confidence interval with model-based variance estimators (Figure \ref{fig:sim_results}.B). In contrast, the highly biased and highly imprecise ETATE estimator had nearly perfect coverage due to its lower precision.

Analyses with an independence correlation structure and model-based variance estimators yielded nominal coverage probability for the CTATE and ETATE estimands when the treatment effect structures were correctly specified, despite misspecification of the correlation structure. However, there was severe under-coverage of the IT effect 95\% confidence intervals for the IT effect estimand when using model-based variance estimators. In sharp contrast, the bias-reduced linearization (CR2) and ``approximate jackknife'' (CR3) robust variance estimators, in analyses with an independence correlation structure, yielded extremely conservative variance estimators as reflected in the extremely low precision. Indeed, such cluster robust variance estimators have previously been observed to be ``extremely conservative in general'' in models with an independence correlation structure \cite{abadie_when_2023}. 

\section{Case Study}
\label{sect:case study}

\begin{figure}[htp]
    \centering
    \includegraphics[width=15cm]{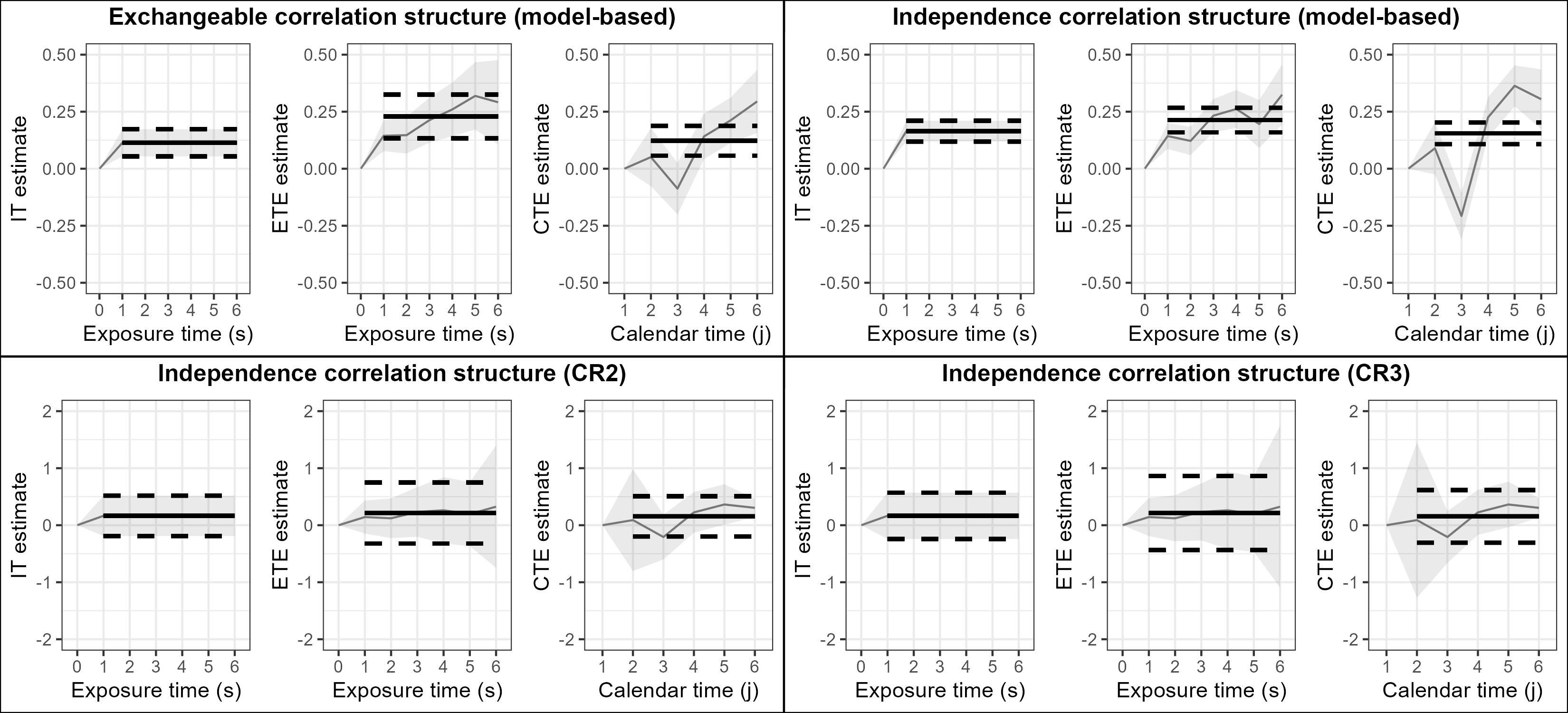}
    \caption{Effect curves over calendar time and exposure time from an Australia disinvestment trial as analyzed with an immediate treatment, exposure time indicators, or calendar time indicators with an exchangeable or independence correlation structure. 95\% confidence intervals (CI's) of each point on the effect curve are shown in gray using model-based variance estimators, and bias-reduced linearization (CR2) and ``approximate jackknife'' (CR3) cluster robust variance estimators for analyses with the independence correlation structure. The solid lines denote the corresponding IT, ETATE, and CTATE estimates, respectively, with analogous dotted lines denoting the 95\% CI's of the IT, ETATE, and CTATE estimates. 
    \newline (*Notably, plots using model-based variance estimators for the 95\% CI's are plotted on a y-axis of $[-0.5, 0.5]$, whereas plots using cluster robust variance estimators for the 95\% CI's are plotted on a y-axis of $[-2,2]$.)}
    \label{fig:disinvestment}
\end{figure}

We reanalyzed case study data collected from a 12 cluster, 7 period SW-CRT examining the impact of removing weekend health services from 12 Australian hospital wards on patient log-length of stay as a continuous outcome variable \cite{haines_impact_2017}. 
We used the models described in Section \ref{sect:models} with an immediate treatment (IT), exposure time indicators (ETI), or calendar time indicators (CTI), and an exchangeable or independence correlation structure. We then plot the corresponding effect curves over exposure time and calendar time in Figure \ref{fig:disinvestment}. Confidence intervals are graphed using the model-based variance estimators in analyses with an exchangeable correlation structure, and with model-based variance estimators, bias-reduced linearization and ``approximate jackknife'' cluster robust variance estimators in analyses with an independence correlation structure (Figure \ref{fig:disinvestment}).

We observe that the IT estimate is similar to the CTATE estimate, but is considerably lower than the ETATE estimate (Figure \ref{fig:disinvestment}). This corresponds with our results in Section \ref{sect:IT_CTATE}, which indicate that the immediate treatment effect estimator does not dramatically differ from the calendar time-averaged treatment effect estimator. However, the immediate treatment effect estimator can differ greatly from the exposure time-averaged treatment effect estimator as previously demonstrated by Kenny et al. \cite{kenny_analysis_2022} and reiterated in Section \ref{sect:IT_ETATE}. 

Analysis with an independence correlation structure yielded IT, ETATE, and CTATE estimates that were closer to one another than the corresponding analyses with an exchangeable correlation structure (Figure \ref{fig:disinvestment}). Still, the resulting estimates between analyses with either correlation structures did not qualitatively differ by much.
\textcolor{black}{We additionally report the AIC and BIC values in analyses with an immediate, exposure time-varying, and calendar time-varying treatment effect structures and an (i.) exchangeable or (ii.) independence correlation structure in Table \ref{tab:case_study_AIC_BIC}. Overall, analyses with a calendar time-varying treatment effect structure had the lowest AIC and BIC values 
(Table \ref{tab:case_study_AIC_BIC}). Still, in a clinical trial, estimands and estimation methods must be pre-specified, and use of AIC or BIC for data-driven model selection will generally not result in valid inference \cite{berk_statistical_2010}, unless it is done so in a framework that explicitly controls type I error.}

With model-based variance estimators, confidence intervals for calendar time-varying treatment effects tend to have a ``bowtie'' shape, with confidence intervals being the most narrow during the midpoint of the trial. Interestingly, this ``bowtie'' shape is not observed with the calendar time-varying treatment effect confidence intervals produced by the cluster robust variances. The cluster robust variance estimators in analyses with an independence correlation structure were extremely conservative, resulting in extremely large confidence intervals, with some resulting cluster robust standard errors being up to 10 times larger than their corresponding model-based standard errors.

We additionally plot the results from mixed effects model analyses with an exchangeable correlation structure and 95\% confidence intervals generated with bias-reduced linearization cluster (CR2) or ``approximate jackknife'' (CR3) robust variance estimators in the Appendix (Appendix \ref{sect:appendix_case_study}). Notably, the overly conservative cluster robust standard errors observed in analyses with an independence correlation structure were not observed in mixed effects model analyses with an exchangeable correlation structure (Appendix \ref{sect:appendix_case_study}).

Based on these results, it is unclear which effect curve is the ``true'' effect curve. \textcolor{black}{While the AIC and BIC appear to prefer models with a calendar time-varying treatment effect, we still generally recommend relying on \textit{a priori} information to make assumptions regarding how the treatment effect may vary over time to inform subsequent analyses.}

\section{Discussion}
\label{sect:discussion}

As discussed in previous literature \cite{wang_how_2024}, the SW-CRT has the unique feature of having two different time scales (exposure time and calendar time), which can yield different time-varying treatment effect estimands. Researchers are generally interested in estimating a time-averaged treatment effect estimand to summarize how the treatment performed over the duration of a study.
However, in this current work, we demonstrate that misspecifation of the treatment effect structure in a mixed effects model analysis can potentially yield severely misleading time-averaged treatment effect estimates.

In Section \ref{sect:IT_ETATE}, we summarize results from Kenny et al. \cite{kenny_analysis_2022}, showing that the immediate treatment effect estimator is the weighted sum of the exposure time-varying treatment effect estimands, with some weights potentially being negative.
In Section \ref{sect:IT_CTATE}, we show that the immediate treatment effect estimator is the weighted sum of the calendar time-varying treatment effect estimands, with all weights being $\geq 0$.
In Section \ref{sect:ETATE_CTATE} (\& \ref{sect:CTATE_ETATE}), we show that the exposure (calendar) time-averaged treatment effect estimator is the weighted sum of the calendar (exposure) time-varying treatment effect estimands, with some weights potentially being negative. 
With such negative weights, the misspecified estimator can potentially converge to an estimate with the opposite sign of the true time-averaged treatment effect estimand.
We further illustrate the potentially severe bias resulting from treatment effect structure misspecification by simulation in Section \ref{sect:simulation}.

While the results in this current work is primarily derived with a mixed effects model with an exchangeable correlation structure \cite{hussey_design_2007}, the results discussed are easily extendable to a mixed effects model with a nested exchangeable correlation structure \cite{hooper_sample_2016,girling_statistical_2016}. The estimand weights in Section \ref{sect:results} are presented as a function of $\gamma$. Accordingly, use of a nested exchangeable mixed effects model simply changes the definition of $\gamma$ from $\gamma=\frac{\tau^{2}_{\alpha}}{\tau^{2}_{\alpha} + \sigma^2_e/K}$ to $\gamma = \frac{\tau^2_\alpha}{\tau^2_\alpha + \tau^2_\omega + \sigma^2_e/K}$, where $\tau^2_\omega$ are the random cluster-period interaction variance terms \cite{hooper_sample_2016,girling_statistical_2016}. 

Notably, in the described conditions where $\gamma=0$ (Figures \ref{fig:ETATE_CTE} \& \ref{fig:CTATE_ETE}), the estimand weights of the misspecified ETATE and CTATE estimators are all $\geq 0$, which should reduce some of the biases resulting from treatment effect structure misspecification. This is equivalent to specifying the analytic model with an independence correlation structure. In our simulations (Section \ref{sect:simulation}), we demonstrated that analyses with an independence correlation structure can yield less biased estimates, even with misspecification of the treatment effect structure. \textcolor{black}{Discussions regarding the analysis of SW-CRTs with models employing an independence correlation structure have previously appeared in Wang et al. \cite{wang_how_2024} and Chen \& Li \cite{chen_model-assisted_2025}. While such analyses can be less efficient, some efficiency can be regained through covariate adjustment if the relevant data is available. Notably, similar models with an independence correlation structure have also been demonstrated to be more robust in the analyses of different CRT designs with informative cluster sizes \cite{wang_two_2022,lee_how_2024}.}

Typically, misspecification of the correlation structure can produce underestimated model-based variance estimators. 
Despite the deliberate misspecification of the correlation structure with an independence correlation structure to reduce bias, the model-based variance estimators for ETATE and CTATE still yielded nominal coverage probabilities when treatment effect structures were correctly specified. However, the IT estimator had under-coverage of the 95\% confidence intervals, as anticipated. Wang et al. \cite{wang_how_2024} and Ouyang et al. \cite{ouyang_maintaining_2024} previously suggested that inference in SW-CRTs can be robust to misspecification of the correlation structure by specifying robust variance estimators. In contrast, we observe that cluster robust variance estimators for time-averaged treatment effects can be overly conservative when used in analyses with an independence correlation structure, corresponding to previous observations by Abadie et al. \cite{abadie_when_2023} in analyses with an independence correlation structure. Noticeably, these extremely conservative cluster robust standard errors were not observed in analyses with an exchangeable correlation structure (Appendix \ref{sect:appendix_case_study}).

\textcolor{black}{Wang et al. \cite{wang_how_2024} also previously demonstrated that provided the treatment effect structure is correctly specified in the analysis of a SW-CRT, with some standard assumptions, mixed effect model analyses will yield consistent estimators for the true time-varying treatment effect estimands regardless of arbitrary model misspecification, which includes adjustment for additional covariates. In a SW-CRT, these covariates are not confounding in expectation due to randomization, assuming the model correctly controls for time. Accordingly, covariate adjustment primarily serves to improve the efficiency. A rigorous demonstration that the described bias results apply to covariate-adjusted working models is outside the scope of the current study.}

\textcolor{black}{In this work, the treatment effect structures are defined with each time-varying treatment effect specified as a discrete fixed variable. Alternatively, a model using a random effect to capture the exposure time-varying treatment effect structure has also been proposed \cite{maleyeff_assessing_2023}. An analogous model with random effects capturing calendar time-varying treatment effects can be explored in future work.} Overall, further work is required to explore more robust methods for analysis of SW-CRTs when the treatment effect structure is unknown.

\subsection{Limitations}

As mentioned, the results derived here are expected to naturally extend to model-based estimators with a nested exchangeable correlation structure. However, it is unclear how the results would extend to models with an exponential decay, unstructured, or other alternative correlation structures \cite{kasza_impact_2019, li_mixed-effects_2021}.

Despite potentially being a key estimand of interest, the ``true" long-term calendar time-varying treatment effect $\xi_J$ is unidentifiable in a typical SW-CRT analytic model that specifies period effect indicators. However, $\xi_J$ can identifiable by either adjusting the study design to include an additional unexposed cluster for the duration of the trial, or by making additional modeling assumptions, such as specifying continuous, rather than discrete, period effects. We do not explore such designs or models in this work.

In current work, we only explore scenarios when there are either exposure or calendar time-varying treatment effects, but not both simultaneously, otherwise referred to as ``saturated'' effects \cite{wang_how_2024}. The properties of a model with such a saturated treatment effect structure are not well understood and outside the scope of this work. Future work can explore and clarify the properties of the saturated model.

\textcolor{black}{While this work has been primarily focused on the standard and complete SW-CRT design, time-varying treatment effects may also be present in other CRT designs. This can include non-standard or incomplete SW-CRT designs, such as the staircase design \cite{grantham_staircase_2024}. Extensions to such designs are beyond the scope of the current work and will be considered in future work.} Lee and Cheung \cite{lee_cluster_2024} point out that the calendar time-varying treatment effects and exposure time-varying treatment effects are inseparable in parallel and parallel with baseline CRT designs. Furthermore, with such parallel designs, the immediate treatment effect (IT), calendar time-averaged treatment effect (CTATE), and exposure time-averaged treatment effect (ETATE) estimators are all equivalent, assuming equal cluster-period cell sizes \cite{lee_cluster_2024}. While this may potentially be interpreted as a drawback for researchers who are interested in separating out the two time-varying treatment effect interactions, it is unclear if such separation is even possible with other CRT designs, such as a SW-CRT. 

\subsection{Conclusions}

Researchers who are interested in applying the SW-CRT study design should be aware of the complicated ways that the treatment effect may vary over time and recognize that misspecification of the treatment effect structure in subsequent analyses can potentially yield severely misleading estimates.
It is crucial for researchers to \textit{a priori} identify whether the treatment effect may vary over exposure time or calendar time when planning a trial, especially those with a stepped-wedge design.

\break
\noindent \textbf{Disclosure statement}

\noindent The authors declare no potential conflicts of interest with respect to the research, authorship, and/or publication of this article.

\noindent \textbf{Funding}

\noindent This article was supported by the National Institute on Deafness and Other Communication Disorders (R01-DC020026) and the National Institute of Dental and Craniofacial Research (U01-OD033247).

\noindent \textbf{Disclaimer}

\noindent The statements presented in this article are solely the responsibility of the authors.

\noindent \textbf{ORCID}

\noindent Kenneth M. Lee (https://orcid.org/0000-0002-0454-4537)

\noindent Elizabeth L. Turner (https://orcid.org/0000-0002-7638-5942)

\noindent Avi Kenny (https://orcid.org/0000-0002-9465-7307)

\noindent \textbf{Data availability statement}

\noindent Data sharing is not applicable to this article as no new data were created or analyzed in this study. R codes for the simulations are hosted on GitHub (https://github.com/Avi-Kenny/Code SW-CalTime).

\break

\bibliographystyle{wileyNJD-AMA}
\bibliography{references}

\clearpage
\begin{table}
\caption{\textcolor{black}{AIC and BIC values with different treatment effect structures in analayses with i.) an exchangeable correlation structure and ii.) an independence correlation structure.}}
\label{tab:case_study_AIC_BIC}
\begin{center}
\bgroup
\def\arraystretch{1.5}
{\color{black}
\begin{tabular}{|c c c c|} 
    \hline
    i.) & \textbf{Treatment Effect Structures (exchangeable)} & \textbf{AIC} & \textbf{BIC}\\
    \hline\hline
    & Immediate & 42952.04 & 43028.08 \\
    \hdashline
    & Exposure time-varying & 42971.96 & 43086.03 \\
    \hdashline
    & Calendar time-varying & 37012.27 & 37109.19 \\
    \hline \hline
    ii.) & \textbf{Treatment Effect Structures (independence)} & \textbf{AIC} & \textbf{BIC}\\
    \hline\hline
    & Immediate & 43772.71 & 43841.15 \\
    \hdashline
    & Exposure time-varying & 43762.17 & 43868.64 \\
    \hdashline
    & Calendar time-varying & 37677.4 & 37766.86 \\
    \hline
\end{tabular}
}
\egroup
\end{center}
\end{table}

\clearpage

\clearpage
\appendix

\section{\textcolor{black}{Appendix: Data-generating process for SW-CRTs with time-varying treatment effects}}
\label{sect:appendix_DGP}
\textcolor{black}{We use a model-based approach to describe the underlying data-generating process of cross-sectional SW-CRTs and to define the time-averaged treatment effect estimands of interest. In a SW-CRT with fixed cluster-period cell sizes, we can specify potential outcomes $Y_{ijk}(q)$ for individual $k \in \{1,...,K\}$ in time period $j \in \{1,...,J\}$ of cluster $i \in \{1,...,I\}$, randomized to sequence $q \in \{1,...,Q\}$, where the treatment in sequence $q$ is introduced during period $j=q+1$ in a SW-CRT. Note that there are typically $Q=J-1$ sequences in a standard complete SW-CRT design, with $I/Q$ clusters equally randomized into each sequence $q$.}

\textcolor{black}{We can then describe a structural model for the generalized conditional expected value of the potential outcome $Y_{ijk}(q)$ for individual $k$ observed in period $j$ of cluster $i$ in sequence $q$. The model is based on the generic structural model outlined in Li et al. \cite{li_mixed-effects_2021} and adapted in Kenny et al. \cite{kenny_analysis_2022}:
\begin{equation}
\label{eq:DGP0}
g(E[Y_{ijk}(q)|\Gamma_j,C_{qijk}])=\Delta_{qij}+\Gamma_j + C_{qijk}
\end{equation}
where $g(.)$ is a link function, $\Delta_{qij}$ is the treatment effect structure, $\Gamma_j$ is the calendar time trend, and $C_{qijk}$ is the cluster-specific, time-specific, and/or individual-specific heterogeneity term which captures the correlation structure of the data \cite{li_mixed-effects_2021}.
Treatment effect structures $\Delta_{qij}$ are unique to each randomization sequence $q$, therefore we do not have to condition on the treatment effect structure $\Delta_{qij}$ in equation \ref{eq:DGP0}.}

\textcolor{black}{Throughout the current article, we focus on models with an identity link function.
Let $Y_{ijk}$ denote the observed outcome for individual $k$ in period $j$ of cluster $i$. 
If we assume that clusters are assigned to sequences randomly (\textit{exchangeability}) and if cluster $i$ is randomized to sequence $q$ (\textit{consistency}), then $E[Y_{ijk}(q)|\Gamma_j,C_{ijk}]$ being the conditional expectation of the potential outcome $Y_{ijk}(q)$ is identified by $E[Y_{ijk}|\Delta_{ij},\Gamma_j,C_{ijk}]$ being the expectation of the observed outcome $Y_{ijk}$ conditioned on the treatment effect structure $\Delta_{ij}$. By conditioning on the treatment effect structure, calendar time trend, and specific departures from the marginal average, we have:
\begin{equation}
\label{eq:DGP2}
E[Y_{ijk}|\Delta_{ij},\Gamma_j,C_{ijk}]=\Delta_{ij}+\Gamma_j+C_{ijk}
\end{equation}
with the assumptions of exchangeability and consistency typically holding in a randomized trial. In equation \ref{eq:DGP2}, the subscripts for sequence $q$ are removed since they are determined by $i$ and $j$ given $\Delta_{ij}$.}

\textcolor{black}{As mentioned in the introduction, most statistical models used to analyze data from stepped wedge designs assume an immediate treatment effect. In the current work, we additionally assume that the treatment effect may vary as a function of exposure time $s_{ij}$ or calendar time period $j$ (but not both). We accordingly rewrite equation \ref{eq:DGP2} as:
\begin{align}
\begin{split}
\label{eq:DGP3.1.appendix}
    E[Y_{ijk}|\dot{X}_{ij},\Gamma_j,C_{ijk}] &= \dot{X}_{ij}\theta + \Gamma_j + C_{ijk} \,,
\end{split} \\
\begin{split}
\label{eq:DGP3.2.appendix}
    E[Y_{ijk}|\ddot{X}_{ij}',\Gamma_j,C_{ijk}] &= \ddot{X}_{ij}'\delta + \Gamma_j + C_{ijk} \,,
\end{split} \\
\begin{split}
\label{eq:DGP3.3.appendix}
    E[Y_{ijk}|\dddot{X}_{ij}',\Gamma_j,C_{ijk}] &= \dddot{X}_{ij}'\xi + \Gamma_j + C_{ijk} \,.
\end{split}
\end{align}
Where the expected outcome $Y_{ijk}$ for individual $k$ observed in period $j$ of cluster $i$ is now conditional on $\dot{X}_{ij}$, $\ddot{X}_{ij}'$, or $\dddot{X}_{ij}'$ being the immediate, exposure time-varying, or calendar time-varying treatment effect structures, respectively.}

\textcolor{black}{Equation \ref{eq:DGP3.1.appendix} represents a general model with an immediate treatment effect structure. Where $q_i$ is the SW-CRT sequence assignment for cluster $i$, then $\dot{X}_{ij} = I(j > q_i)$ is the indicator ($=1$ if assigned to treatment, $=0$ if assigned to control, when a given period $j$ is higher than that of a sequence $q$) for the (single) immediate treatment effect $\theta$ for the individuals observed during period $j$ of cluster $i$ (equation \ref{eq:DGP3.1.appendix}).}

\textcolor{black}{Equation \ref{eq:DGP3.2.appendix} represents a general model with an exposure time-varying treatment effect structure. For simplicity of notation, we use the subscript $s\in(1,2,...,J-1)$ to represent the treatment exposure time, with $s_{ij}$ corresponding to that of cluster $i$ at time $j$. Accordingly, $\ddot{X}_{ij}'=\left(I(s_{ij}=1), I(s_{ij}=2),...,I(s_{ij}=J-1)\right)$ is the 1 by $J-1$ row vector of indicators corresponding to $\delta=(\delta_1,\delta_2,...,\delta_{J-1})'$ as the $J-1$ by 1 column vector of the different exposure time-varying treatment effects $\delta_s$ for individuals observed during period $j$ of cluster $i$.}

\textcolor{black}{Finally, equation \ref{eq:DGP3.3.appendix} represents a general model with a calendar time-varying treatment effect structure. Accordingly, $\dddot{X}_{ij}' = \left(I(j=2 \ \& \ j>q_i),I(j=3 \ \& \ j>q_i),...,I(j=J \ \& \ j>q_i)\right)$ is the 1 by $J-1$ row vector of indicators corresponding to $\xi=(\xi_2,\xi_3,...,\xi_{J})'$ as the $J-1$ by 1 column vector of the different calendar time-varying treatment effects $\xi_j$ for individuals observed during period $j$ of cluster $i$.}

\textcolor{black}{Notably, equation \ref{eq:DGP3.1.appendix} is a special case of equations \ref{eq:DGP3.2.appendix} and \ref{eq:DGP3.3.appendix}, where $\delta_s=\theta \, \forall \, s$ and $\xi_j=\theta \, \forall \, j$, respectively.}

\break
\section{Appendix: Proof of Theorem \ref{theorem:IT_CTE}}
\label{sect:appendix_derivation_IT_CTE}

Derivation of the IT estimator when the treatment effect varies with calendar time.

Assuming the true underlying marginal model has calendar time-varying treatment effects:
\[E[\Bar{Y}_{ij}|\dddot{X}_{ij}', P_j]=\dddot{X}_{ij}'\xi+P_j\phi \,,\]
we can use equation \ref{eq:IT} to demonstrate that the expected value of the immediate treatment effect is:
\[E[\widehat{IT}|\dddot{X}_{ij}', P_j]= 
\frac{12(1+\gamma Q)}{Q(Q+1)(\gamma Q^2+2Q-\gamma Q-2)}\sum_{j=1}^{J}\sum_{q=1}^{Q}\left[Q(I(j>q))+1-j+\frac{\gamma Q(2q-Q-1)}{2(1+\gamma Q)}\right]E[\Bar{Y}_{qj}|\dddot{X}_{ij}', P_j]\]

\noindent summed across sequences $q \in \{1,...,Q\}$ and periods $j \in \{1,...,J\}$. Since (1.) $\xi_1=\xi_J=0$, and (2.) rewriting $\dddot{X}_{ij}'\xi = I(j>q)\xi_j$ (with $I(j>q)$ referring to an indicator for whether the index for a given period $j$ is higher than that of a given sequence $q$), this can be written as:

\[E[\widehat{IT}|\dddot{X}_{ij}', P_j]= 
M\sum_{j=2}^{J-1}\sum_{q=1}^{Q}\left[Q(I(j>q))+1-j+\frac{\gamma Q(2q-Q-1)}{2(1+\gamma Q)}\right] (I(j>q)\xi_j + P_j \phi)) \,, \]

\noindent where
\[
M \equiv \frac{12(1+\gamma Q)}{Q(Q+1)(\gamma Q^2+2Q-\gamma Q-2)} \,.
\]

\noindent In Kenny et al. \cite{kenny_analysis_2022}, the following equality was shown to hold:

\[ \sum_{j=1}^{J}\sum_{q=1}^{Q}\left[Q(I(j>q))+1-j+\frac{\gamma Q(2q-Q-1)}{2(1+\gamma Q)}\right] P_j \phi = 0 \,. \]

\noindent Therefore, we can write:
\begin{align*}
\begin{split}
    E[\widehat{IT}|\dddot{X}_{ij}', P_j]
    &= M\sum_{j=2}^{J-1}\sum_{q=1}^{Q}\left[Q(I(j>q))+1-j+\frac{\gamma Q(2q-Q-1)}{2(1+\gamma Q)}\right] I(j>q)\xi_j
\end{split} \\
\begin{split}
    &= M\sum_{j=2}^{J-1}\sum_{q=1}^{j-1}\left[Q+1-j+\frac{\gamma Q(2q-Q-1)}{2(1+\gamma Q)}\right] \xi_j
\end{split} \\
\begin{split}
    &= M\sum_{j=2}^{J-1}\left[(j-1)(Q+1-j)+\frac{\gamma Q}{2(1+\gamma Q)}\sum_{q=1}^{j-1}(2q-Q-1)\right]\xi_j \, .
\end{split}
\end{align*}

\noindent Since $\sum_{q=1}^{j-1}q=\frac{j(j-1)}{2}$, it holds that:

\[E[\widehat{IT}|\dddot{X}_{ij}', P_j] = M\sum_{j=2}^{J-1}\left[(j-1)(Q+1-j)+\frac{(\gamma Q)(j-1)}{2(1+\gamma Q)}(-Q-1+j)\right]\xi_j.\]

Altogether, the immediate treatment estimator is in expectation a weighted average of the calendar time-varying treatment effect estimands $\xi_j$:

\begin{align*}
\begin{split}
    E[\widehat{IT}|\dddot{X}_{ij}', P_j]
    &= \frac{12(1+\gamma Q)}{Q(Q+1)(Q-1)(\gamma Q+2)}\sum_{j=2}^{J-1}\left[(j-1)(Q+1-j)\left(\frac{2+ \gamma Q}{2+2\gamma Q}\right)\right]\xi_j
\end{split} \\
\begin{split}
    &= \sum_{j=2}^{J-1} \frac{6(j-1)(Q+1-j)}{Q(Q+1)(Q-1)} \xi_j \,.
\end{split}
\end{align*}

\noindent This concludes the proof. $\square$

\textcolor{black}{Notably, Theorem \ref{theorem:IT_CTE} and the rest of the subsequent derivations are done with generalized least squares, assuming the true variance components $\tau^2_\alpha$, $\sigma^2_e$, and $\gamma = \frac{\tau^2_\alpha}{\tau^2_\alpha + \sigma^2_e/K}$ are known. If implemented with feasible generalized least squares, using the estimated variance components $\hat{\tau}^2_\alpha$, $\hat{\sigma}^2_e$, and $\hat{\gamma} = \frac{\hat{\tau}^2_\alpha}{\hat{\tau}^2_\alpha + \hat{\sigma}^2_e/K}$, the results can still hold across values of the estimated $\hat{\gamma}$ instead of the true $\gamma$, as shown here.}

\textcolor{black}{Still, Theorem \ref{theorem:IT_CTE} holds regardless of whether generalized least squares or feasible generalized least squares was used, as the immediate treatment effect estimator is shown above to be a weighted sum of the true calendar time-varying treatment effect estimands, independent of the variance components, known or unknown.}

\break
\section{Appendix: Behavior of the ETATE estimator with a true underlying calendar time-varying treatment effect structure}
\label{sect:appendix_derivation_ETATE_CTE}

Derivation of the ETATE estimator when the treatment effect varies with calendar time

The ETI mixed effects model (equation \ref{eq:ETImodel}):
\[
Y_{ijk}=\ddot{X}_{ij}'\delta+P_j\phi+\alpha_{i}+\epsilon_{ijk} \,,
\]

\noindent can be re-written with matrix notation using the following augmented design matrix $\ddot{Z}=(\ddot{X}_{ij}'|P_j)$, and augmented vector of coefficients $\beta_{ETI}=(\delta'|\phi')'$:
\[
    Y=\ddot{Z}\beta_{ETI}+v \,.
\]
We can specify this above formula with cluster-period means, with $Y$ being the vector of cluster-period mean outcomes $\bar{Y}_{ij}$ over individuals in period $j \in \{1,...,J\}$ of cluster $i \in \{1,...,I\}$, and $v$ being the vector of heterogeneity terms $v_{ij}=\alpha_i+\epsilon_{ij}$ on the cluster-period cell level, where $v \overset{iid}{\sim} N(0,V)$. In such a mixed effects model with an exchangeable correlation structure, $V=\mathbb{I}_I \bigotimes R_i$ is the $IJ$ by $IJ$ block diagonal variance-covariance matrix of $Y$ (where $\mathbb{I}_I$ is an $I$ by $I$ dimension identity matrix) and:
\[R_i = Var(\bar{Y}_{ij})(\mathbb{I}_J(1-\gamma) + \mathbb{J}_j(\gamma))\]
(where $\mathbb{I}_J$ and $\mathbb{J}_J$ are $J$ by $J$ dimension matrices, representing the identity matrix and a matrix of ones, respectively), and where $\bigotimes$ represents the Kronecker product. 
With an exchangeable correlation structure and fixed cluster-period sizes $K$, $Var(\bar{Y}_{ij}) = Var(v_{ij}) = \tau^{2}_{\alpha} + \sigma^2_e/K$ and cancels out of the generalized least squares (GLS) point estimator.

The GLS point estimator for the ETI mixed effects model can accordingly be written as:
\[
\widehat{\beta}_{ETI}=(\ddot{Z}'V^{-1}\ddot{Z})^{-1}\ddot{Z}'V^{-1}Y \,.
\]

\subsection{Derivation of the ETATE estimator when the treatment effect varies with calendar time in a standard $Q$ sequence, $J$ period SW-CRT design}
\label{sect:appendix_derivation_ETATE_CTE_1}

Here, we outline how the ETATE estimator can be written as a weighted average of the CTI estimands in a simple and standard SW-CRT design with $I$ clusters and $J$ periods, where there is $I/Q=1$ cluster in each randomization sequence $q \in \{1,...,Q\}$ and $I=J-1$. This yields an equivalent point estimator to scenarios with $Q$ sequences with equal allocation of $I/Q>1$ clusters randomized into each sequence $q$.

We must first derive the ETI point estimators $\hat{\delta}_s$ where $s \in \{1,..,J-1\}$. To do so, we first derive $(\ddot{Z}'V^{-1}\ddot{Z})$
 with block matrices, such that:
\[
(\ddot{Z}'V^{-1}\ddot{Z})=\frac{1}{\sigma^2+\tau_{\alpha}^2}
\begin{pmatrix}
\Omega_{11} & \Omega_{12}\\
\Omega_{21} & \Omega_{22}
\end{pmatrix}
\]
where $\sigma^2=\frac{\sigma^2_e}{K}$. Here, $\Omega_{11}$ is the top-left $(J-1)$ by $(J-1)$ block matrix corresponding with the ETI effects $\delta_s$, and $\Omega_{22}$ is the bottom-right $J$ by $J$ matrix corresponding with the time period effects $\phi_j$.
Crucially, $(\ddot{Z}'V^{-1}\ddot{Z})$ using information from all observed cluster-period cells is a positive-definite matrix; hence we can define:
\[
(\ddot{Z}'V^{-1}\ddot{Z})^{-1}=(\sigma^2+\tau_{\alpha}^2)
\begin{pmatrix}
\Omega_{11} & \Omega_{12}\\
\Omega_{21} & \Omega_{22}
\end{pmatrix}^{-1} \,.
\]

The first $J-1$ rows of $(\ddot{Z}'V^{-1}\ddot{Z})^{-1}$ corresponding to the ETI effects is then equal to:
\[
(\ddot{Z}'V^{-1}\ddot{Z})^{-1}_{[\delta_1,...,\delta_{J-1}]} = (\sigma^2+\tau^2_\alpha)
\begin{pmatrix}
(\Omega_{11}-\Omega_{12}\Omega_{22}^{-1}\Omega_{21})^{-1} & -(\Omega_{11}-\Omega_{12}\Omega_{22}^{-1}\Omega_{21})^{-1}\Omega_{12}\Omega_{22}^{-1}
\end{pmatrix} \,.
\]
Subsequently, $(\ddot{Z}'V^{-1}\ddot{Z})^{-1}_{[\delta_1,...,\delta_{J-1}]}\ddot{Z}'V^{-1}$ is the $(J-1)$ by $J(J-1)$ matrix:
\[
(\ddot{Z}'V^{-1}\ddot{Z})^{-1}_{[\delta_1,...,\delta_3]}\ddot{Z}'V^{-1} =
\begin{pmatrix}
\lambda_{[\delta_1]}\\
\vdots\\
\lambda_{[\delta_{J-1}]}
\end{pmatrix}
\]
where $\lambda_{[\delta_s]}$ is the row vector corresponding to coefficient $\delta_s$ in $(\ddot{Z}'V^{-1}\ddot{Z})^{-1}_{[\delta_1,...,\delta_{J-1}]}\ddot{Z}'V^{-1}$, with $J(J-1)$ entries of $\lambda_{[\delta_s]ij}$ corresponding to cluster-period mean observations from period $j \in \{1,...,J\}$ of cluster $i \in \{1,...,I\}$, with $I=J-1$.

Finally, $\hat{\delta}$, the resulting vector of the ETI estimators is:
\[
\hat{\delta}=(\ddot{Z}'V^{-1}\ddot{Z})^{-1}_{[\delta_1,...,\delta_{J-1}]}\ddot{Z}'V^{-1}Y \,.
\]
Recall that $Y$ is the vector of cluster-period cell means $\bar{Y}_{ij}$. As a result, we can write each $\hat{\delta}_s$ as:
\[
\hat{\delta}_s =\sum_{i=1}^{I}\sum_{j=1}^{J}\lambda_{[\delta_s]ij}\bar{Y}_{ij} \,.
\]

Assuming the underlying marginal model has calendar time-varying treatment effects, such that:
\[
E[\Bar{Y}_{ij}|\dddot{X}_{ij}', P_j]=\dddot{X}_{ij}'\xi+P_j\phi
\]
is true, the conditional expectation of a given ETI estimator $\hat{\delta}_s$, given covariates $\dddot{X}_{ij}'$ for the true CTI treatment effect structure is then:
\begin{align*}
\begin{split}
E[\hat{\delta}_s|\dddot{X}_{ij}',P_j]=\sum_{i=1}^{I}\sum_{j=1}^{J}\lambda_{[\delta_s]ij}E[\bar{Y}_{ij}|\dddot{X}_{ij}',P_j]
\end{split} \\
\begin{split}
&=\sum_{i=1}^{I}\sum_{j=1}^{J}\lambda_{[\delta_s]ij}(\dddot{X}_{ij}'\xi+P_j\phi) \,.
\end{split}
\end{align*}

\noindent In Kenny et al. \cite{kenny_analysis_2022}, the following equality was shown to hold:

\[ \sum_{i=1}^{I}\sum_{j=1}^{J}\lambda_{[\delta_s]ij} (P_j \phi) = 0 \,. \]

\noindent Furthermore, we assume with a true underlying calendar time-varying treatment effect, (1.) $\xi_1 = \xi_J = 0$, because the first and last periods ($j=1$ and $j=J$) have no within-period variation in treatment effect (being unexposed and all exposed to treatment, respectively). As a result, the observations from these periods do not contribute to $E[\hat{\delta}_s|\dddot{X}_{ij}',P_j]$. Therefore, we sum over periods $j \in [2,J-1]$, and observe that within these periods, only clusters $i \in [1,j-1]$ receive the corresponding calendar-time varying treatment $\xi_j$. Additionally, rewriting (2.) $\dddot{X}_{ij}'\xi = I(j>i)\xi_j$ (with $I(j>i)$ referring to an indicator for whether the index for a given period $j$ is higher than that of a given sequence $q$), altogether:
\begin{align*}
\begin{split}
    E[\hat{\delta}_s|\dddot{X}_{ij}',P_j]=\sum_{i=1}^{I}\sum_{j=2}^{J-1}\lambda_{[\delta_s]ij} (I(j>i)\xi_j)) 
\end{split} \\
\begin{split}
    &= \sum_{j=2}^{J-1}\sum_{i=1}^{j-1}\lambda_{[\delta_s]ij}\xi_j \,.
\end{split}
\end{align*}

\noindent This expression is equivalent for scenarios where multiple $I/Q>1$ clusters are equally allocated to each sequence $q$, where we instead have sequence-specific weights $\lambda_{[\delta_s]qj}$ summed over sequences $q \in \{1, ..., j-1\}$ (rather than cluster-specific weights summed over clusters $i$). Accordingly:
\[
E[\hat{\delta}_s|\dddot{X}_{ij}',P_j]= \sum_{j=2}^{J-1}\sum_{q=1}^{j-1}\lambda_{[\delta_s]qj}\xi_j \,.
\]

With all the individual ETI estimators $\hat{\delta}_s$ being shown to be weighted sums of the CTI estimands, we can then demonstrate that the ETATE estimator can also be written as a weighted sum of the CTI estimands:

\begin{align*}
\begin{split}
    E[\widehat{ETATE}|\dddot{X}_{ij}',P_j] = \frac{\sum_{s}^{J-1}E[\hat{\delta}_s|\dddot{X}_{ij}',P_j]}{J-1} 
\end{split} \\
\begin{split}
    &=\frac{\sum_{s=1}^{J-1}\sum_{j=2}^{J-1}\sum_{q=1}^{j-1}\lambda_{[\delta_s]qj}\xi_j}{J-1}
\end{split} \\
\begin{split}
    &=\sum_{j=2}^{J-1}w_3(Q,\gamma,j)\xi_j
\end{split}
\end{align*}

\noindent with estimand weights $w_3(Q,\gamma,j)=\frac{\sum_{s=1}^{J-1}\sum_{q=1}^{j-1}\lambda_{[\delta_s]qj}}{J-1}$, total number of sequences $Q=J-1$, and $\gamma=\frac{\tau^{2}_{\alpha}}{\tau^{2}_{\alpha} + \sigma^2_e/K}$ assuming equal cluster-period size $K$ and an exchangeable correlation structure.

Altogether, we demonstrate that the ETATE estimator can be written as a weighted average of the CTI estimands (with weights $w_3(Q,\gamma,j)$) in a simple and standard $Q$ sequence and $J$ period SW-CRT design, where there are $I/Q$ clusters in each randomization sequence $q$. $\square$

\subsection{Derivation of the ETATE estimator when the treatment effect varies with calendar time in a 3 sequence, 4 period SW-CRT design}
\label{sect:appendix_derivation_ETATE_CTE_2}

We demonstrate the derivation outlined in Appendix Section \ref{sect:appendix_derivation_ETATE_CTE_1} with a $Q=3$ sequence, $J=4$ period SW-CRT design. Again, we start the derivation assuming $I$ clusters, with $I/Q=1$ cluster equally allocated into each sequence $q$.
Accordingly:
\begin{align*}
\begin{split}
    E[\hat{\delta}_s|\dddot{X}_{ij}',P_j]=\sum_{j=2}^{3}\sum_{i=1}^{j-1}\lambda_{[\delta_s]ij}\xi_j
\end{split} \\
\begin{split}
    &=\lambda_{[\delta_s]12}\xi_2 + \lambda_{[\delta_s]13}\xi_3 + \lambda_{[\delta_s]23}\xi_3
\end{split}
\end{align*}

Altogether, we can derive and demonstrate that the expected value of the ETATE estimator, given that the true treatment effect structure has calendar time-varying treatment effects is equal to:
\begin{align*}
\begin{split}
    E[\widehat{ETATE}|\dddot{X}_{ij}',P_j] = \frac{\sum_{s=1}^{J-1}E[\hat{\delta}_s|\dddot{X}_{ij}',P_j]}{J-1} 
\end{split} \\
\begin{split}
    &= \frac{E[\hat{\delta}_1+\hat{\delta}_2+\hat{\delta}_3|\dddot{X}_{ij}',P_j]}{3}
\end{split} \\
\begin{split}
    &=\frac{\sum_{s=1}^{3}(\lambda_{[\delta_s]12})\xi_2 + \sum_{s=1}^{3}(\lambda_{[\delta_s]13}+\lambda_{[\delta_s]23})\xi_3}{3}
\end{split} \\
\begin{split}
    &=w_3(Q,\gamma,2)\xi_2 + w_3(Q,\gamma,3)\xi_3
\end{split}
\end{align*}
where $w_3(Q,\gamma,j)$ are the CTI estimand-specific weights. Analytically solving for these weights $w_3(Q,\gamma,j)$ results in:
\[
    E[\widehat{ETATE}|\dddot{X}_{ij}',P_j] =\frac{-9\gamma^2+30\gamma+12}{2(9\gamma^2+39\gamma+13)}\xi_2+\frac{27\gamma^2+48\gamma+14}{2(9\gamma^2+39\gamma+13)}\xi_3
\]
where $\gamma=\frac{\tau^{2}_{\alpha}}{\tau^{2}_{\alpha} + \sigma^2_e/K}$ with equal cluster-period size $K$.

\break
\section{Appendix: Behavior of the CTATE estimator with a true underlying exposure time-varying treatment effect structure}
\label{sect:appendix_derivation_CTATE_ETE}

Derivation of the CTATE estimator when the treatment effect varies with exposure time.

The CTI mixed effects model:
\[
Y_{ijk}=\dddot{X}_{ij}'\xi+P_j\phi+\alpha_{i}+\epsilon_{ijk} \,,
\]

\noindent can be re-written with matrix notation, using the following augmented design matrix $\dddot{Z}=(\dddot{X}_{ij}'|P_j)$, and augmented vector of coefficients $\beta_{CTI}=(\xi'|\phi')'$:
\[
    Y=\dddot{Z}\beta_{CTI}+v_{ij} \,.
\]
We can specify this with cluster-period means, with $Y$ being the vector of cluster-period mean outcomes $\bar{Y}_{ij}$ and $v$ being the vector of heterogeneity terms $v_{ij}=\alpha_i+\epsilon_{ij}$, where $v \overset{iid}{\sim} N(0,V)$. IN such a mixed effects model with an exchangeable correlation structure, $V=\mathbb{I}_I \bigotimes R_i$ is the $IJ$ by $IJ$ block diagonal variance-covariance matrix of $Y$ (where $\mathbb{I}_I$ is an $I$ by $I$ dimension identity matrix) and:
\[R_i = Var(\bar{Y}_{ij})(\mathbb{I}_J(1-\gamma) + \mathbb{J}_j(\gamma))\]
(where $\mathbb{I}_J$ and $\mathbb{J}_J$ are $J$ by $J$ dimension matrices, representing the identity matrix and a matrix of ones, respectively), and where $\bigotimes$ represents the Kronecker product. With an exchangeable correlation structure and fixed cluster-period sizes $K$, $Var(\bar{Y}_{ij}) = Var(v_{ij}) = \tau^{2}_{\alpha} + \sigma^2_e/K$ and cancels out of the generalized least squares (GLS) point estimator.

The GLS point estimators for the CTI mixed effects model can accordingly be written as:
\[
\widehat{\beta}_{CTI}=(\dddot{Z}'V^{-1}\dddot{Z})^{-1}\dddot{Z}'V^{-1}Y \,.
\]

\subsection{Derivation of the CTATE estimator when the treatment effect varies with exposure time in a standard $Q$ sequence, $J$ period SW-CRT design}
\label{sect:appendix_derivation_CTATE_ETE_1}

Here, we outline how the CTATE estimator can be written as a weighted average of the ETI estimands in a simple and standard SW-CRT design with $I$ clusters and $J$ periods, where there is $I/Q=1$ cluster in each randomization sequence $q \in \{1,...,Q\}$ and $I=J-1$. This yields an equivalent estimator to scenarios with $Q$ sequences with equal allocation of $I/Q>1$ clusters randomized into each sequence $q$.

We must first derive the CTI point estimators $\hat{\xi}_j$ where $j \in \{2,...,J-1\}$. Notably, $\hat{\xi}_J$ and $\hat{\phi}_J$ from the final period $j=J$ is unidentifiable. Importantly, observations from period $J$ are accordingly excluded from this analysis due to all clusters being exposed to the treatment during this period. To derive these CTI point estimators, we first derive $(\dddot{Z}'V^{-1}\dddot{Z})$
 with block matrices, such that:
\[
(\dddot{Z}'V^{-1}\dddot{Z})=\frac{1}{\sigma^2+\tau_{\alpha}^2}
\begin{pmatrix}
\Omega_{11} & \Omega_{12}\\
\Omega_{21} & \Omega_{22}
\end{pmatrix}
\]
where $\sigma^2=\frac{\sigma^2_e}{K}$. Here, $\Omega_{11}$ is the top-left $(J-2)$ by $(J-2)$ block matrix corresponding with the CTI effects, and $\Omega_{22}$ is the bottom-right $(J-1)$ by $(J-1)$ matrix corresponding with the time period effects. 
Crucially, $(\dddot{Z}'V^{-1}\dddot{Z})$ using information from observed cluster-period cells where $j \neq J$ is a positive-definite matrix; hence we can define:
\[
(\dddot{Z}'V^{-1}\dddot{Z})^{-1}=(\sigma^2+\tau_{\alpha}^2)
\begin{pmatrix}
\Omega_{11} & \Omega_{12}\\
\Omega_{21} & \Omega_{22}
\end{pmatrix}^{-1}
\]

The first J-2 rows of $(\dddot{Z}'V^{-1}\dddot{Z})^{-1}$ corresponding to the CTI effects is equal to:
\[
(\dddot{Z}'V^{-1}\dddot{Z})^{-1}_{[\xi_2,...,\xi_{J-1}]} = (\sigma^2+\tau^2_\alpha)
\begin{pmatrix}
(\Omega_{11}-\Omega_{12}\Omega_{22}^{-1}\Omega_{21})^{-1} & -(\Omega_{11}-\Omega_{12}\Omega_{22}^{-1}\Omega_{21})^{-1}\Omega_{12}\Omega_{22}^{-1}
\end{pmatrix}
\]
Subsequently, $(\dddot{Z}'V^{-1}\dddot{Z})^{-1}_{[\xi_2,...,\xi_{J-1}]}\dddot{Z}'V^{-1}$ is the $(J-2)$ by $(J-1)^2$ matrix:
\[
(\dddot{Z}'V^{-1}\dddot{Z})^{-1}_{[\xi_2,...,\xi_{J-1}]}\dddot{Z}'V^{-1} =
\begin{pmatrix}
\lambda_{[\xi_2]}\\
\vdots \\
\lambda_{[\xi_{J-1}]}
\end{pmatrix}
\]
where $\lambda_{[\xi_j]}$ is the row vector corresponding to coefficient $\xi_j$ in $(\dddot{Z}'V^{-1}\dddot{Z})^{-1}_{[\xi_2,...,\xi_{J-1}]}\dddot{Z}'V^{-1}$, with $(J-1)^2$ entries $\lambda_{[\xi_j]ij}$ corresponding to cluster-period mean observations from period $j\in\{1,...,J-1\}$ of cluster $i\in\{1,...,I\}$ where $I=J-1$.

Finally, $\hat{\xi}$, the resulting vector of the CTI estimators is:
\[
\hat{\xi}=(\dddot{Z}'V^{-1}\dddot{Z})^{-1}_{[\xi_2,...,\xi_{J-1}]}\dddot{Z}'V^{-1}Y \,.
\]
Recall that $Y$ is the vector of cluster-period cell means $\bar{Y}_{ij}$. As a result, we can write each $\hat{\xi}_{j=c}$ $\forall c \in \{2,...,J-1\}$ as:
\[
\hat{\xi}_{j=c} =\sum_{i=1}^{I}\sum_{j=1}^{J-1}\lambda_{[\xi_{j=c}]ij}\bar{Y}_{ij} \,.
\]

Assuming the underlying marginal model has exposure time-varying treatment effects, such that:
\[
E[\Bar{Y}_{ij}|\ddot{X}_{ij}', P_j]=\ddot{X}_{ij}'\delta+P_j\phi
\]
is true, the conditional expectation of the CTI estimator $\hat{\xi}_{j=c}$, given covariates $\ddot{X}_{ij}'$ for the true ETI treatment effect structure is then:
\begin{align*}
\begin{split}
    E[\hat{\xi}_{j=c}|\ddot{X}_{ij}',P_j] =\sum_{i=1}^{I}\sum_{j=1}^{J-1}\lambda_{[\xi_{j=c}]ij}E[\bar{Y}_{ij}|\ddot{X}_{ij}',P_j] 
\end{split} \\
\begin{split}
    &=\sum_{i=1}^{I}\sum_{j=1}^{J-1}\lambda_{[\xi_{j=c}]ij}(\ddot{X}_{ij}'\delta+P_j\phi) \,.
\end{split}
\end{align*}

\noindent In Kenny et al. \cite{kenny_analysis_2022}, the following equality was shown to hold:

\[ \sum_{i=1}^{I}\sum_{j=1}^{J-1}\lambda_{[\xi_{j=c}]ij}(P_j\phi)= 0 \]

\noindent $\forall c \in \{2,...,J-1\}$. Therefore, (1.) we only sum over periods $j \in [2,J-1]$ that have some cluster-period cells receiving the control and others receiving the treatment. Subsequently, we observe that within these periods, only clusters $i \in [1,j-1]$ receive the treatment. Furthermore, (2.) cluster-period cells receiving the treatment during period $j$ of cluster $i$ will receive the exposure-time varying treatment $\delta_{j-1}$. Therefore, we can rewrite $\ddot{X}_{ij}'\delta = I(j>i)\delta_{j-i}$ (with $I(j>i)$ referring to an indicator for whether the index for a given period $j$ is higher than that of a given sequence $q$). Altogether:
\begin{align*}
\begin{split}
    E[\hat{\xi}_{j=c}|\ddot{X}_{ij}',P_j]=\sum_{i=1}^{I}\sum_{j=2}^{J-1}\lambda_{[\xi_{j=c}]ij}(I(j>i)\delta_{j-i})
\end{split} \\
\begin{split}
    &=\sum_{j=2}^{J-1}\sum_{i=1}^{j-1}\lambda_{[\xi_{j=c}]ij}\delta_{j-i}
\end{split}
\end{align*}

\noindent This is equivalent for scenarios where multiple $I/Q>1$ clusters are equally allocated to each sequence $q$, where we instead have sequence-specific weights $\lambda_{[\xi_j]qj}$, summed over sequences $q \in \{1, ..., j-1\}$ (rather than cluster-specific weights and summed over clusters $i$), with observations in sequence $q$ and period $j$ receiving ETI effect $\delta_{j-q}$ given that $j>q$. Accordingly:
\[
E[\hat{\xi}_{j=c}|\ddot{X}_{ij}',P_j]=\sum_{j=2}^{J-1}\sum_{q=1}^{j-1}\lambda_{[\xi_{j=c}]qj}\delta_{j-q}
\]

With all the individual CTI estimators $\hat{\xi}_j$ being shown to be weighted sums of the ETI estimands, we can then demonstrate that the CTATE estimator can also be written as a weighted sum of the ETI estimands:
\begin{align*}
\begin{split}
    E[\widehat{CTATE}|\ddot{X}_{ij}',P_j] = \frac{\sum_{c=2}^{J-1}E[\hat{\xi}_{j=c}|\ddot{X}_{ij}',P_j]}{J-2} 
\end{split} \\
\begin{split}
    &= \frac{\sum_{c=2}^{J-1}\sum_{j=2}^{J-1}\sum_{q=1}^{j-1}\lambda_{[\xi_{j=c}]qj}\delta_{j-q}}{J-2} \,.
\end{split}
\end{align*}
Writing the expression in terms of exposure time $s$, we have:
\begin{align*}
\begin{split}
    E[\widehat{CTATE}|\ddot{X}_{ij}',P_j] =\frac{\sum_{s=1}^{J-2} (\sum_{c=2}^{J-1}\sum_{j=2}^{J-1} \sum_{q=1}^{J-1} I(j-q=s)\lambda_{[\xi_{j=c}]qj}) \delta_{s}}{J-2}
\end{split} \\
\begin{split}
    &=\sum_{s=1}^{J-2} w_{4}(Q,\gamma,s) \delta_s
\end{split}
\end{align*}
with estimand weights $w_{4}(Q,\gamma,s)=\frac{\sum_{c=2}^{J-1}\sum_{j=2}^{J-1} \sum_{q=1}^{J-1} I(j-q=s)\lambda_{[\xi_{j=c}]qj}}{J-2}$ (with $I(j-q=s)$ referring to an indicator for indices where the difference between the given period $j$ and sequence $q$ is equal to the given exposure time $s$, $j-q=s$), total number of sequences $Q=J-1$, and $\gamma=\frac{\tau^{2}_{\alpha}}{\tau^{2}_{\alpha} + \sigma^2_e/K}$ assuming equal cluster-period size $K$ and an exchangeable correlation structure.

Altogether, we demonstrate that the CTATE estimator can be written as a weighted average of the ETI estimands (with weights $w_{4}(Q,\gamma,s)$) in a simple and standard $Q$ sequence and $J$ period SW-CRT design, where there are $I/Q$ clusters in each randomization sequence $q$. $\square$

\subsection{Derivation of the CTATE estimator when the treatment effect varies with exposure time in a 3 sequence, 4 period SW-CRT design}
\label{sect:appendix_derivation_CTATE_ETE_2}

We demonstrate the derivation outlined in Appendix Section \ref{sect:appendix_derivation_CTATE_ETE_1} with a $Q=3$ sequence, $J=4$ period SW-CRT design. Again, we start the derivation assuming $I$ clusters with $I/Q=1$ cluster equally allocated into each sequence $q$. Accordingly:
\begin{align*}
\begin{split}
    E[\hat{\xi}_{j=c}|\ddot{X}_{ij}',P_j]=\sum_{j=2}^{3}\sum_{i=1}^{j-1}\lambda_{[\xi_{j=c}]ij}\delta_{j-i}
\end{split} \\
\begin{split}
    &=\lambda_{[\xi_{j=c}]12}\delta_1 + \lambda_{[\xi_{j=c}]13}\delta_2 + \lambda_{[\xi_{j=c}]23}\delta_1
\end{split}
\end{align*}

Altogether, we can derive and demonstrate that the expected value of the CTATE estimator, given that the true treatment effect structure has exposure time-varying treatment effects is equal to:
\begin{align*}
\begin{split}
    E[\widehat{CTATE}|\ddot{X}_{ij}',P_j] = \frac{\sum_{c=2}^{J-1}E[\hat{\xi}_{j=c}|\ddot{X}_{ij}',P_j]}{J-2} 
\end{split} \\
\begin{split}
    &= \frac{E[\hat{\xi}_2+\hat{\xi}_3|\ddot{X}_{ij}',P_j]}{2}
\end{split} \\
\begin{split}
    &=\frac{\sum_{j=2}^{3}(\lambda_{[\xi_{j=c}]12}+\lambda_{[\xi_{j=c}]23})\delta_1 + \sum_{j=2}^{3}(\lambda_{[\xi_{j=c}]13})\delta_2}{2}
\end{split} \\
\begin{split}
    &=w_{4}(Q,\gamma,1) \delta_1 + w_{4}(Q,\gamma,2) \delta_2
\end{split}
\end{align*}
where $w_{4}(Q,\gamma,s)$ are the ETI estimand-specific weights corresponding to exposure time $s$. Analytically solving for these weights $w_{4}(Q,\gamma,s)$ results in:
\[
    E[\widehat{CTATE}|\ddot{X}_{ij}',P_j] =\frac{9\gamma^2+15\gamma+6}{2(3\gamma^2+8\gamma+4)}\delta_1+\frac{-3\gamma^2+\gamma+2}{2(3\gamma^2+8\gamma+4)}\delta_2
\]
where $\gamma=\frac{\tau^{2}_{\alpha}}{\tau^{2}_{\alpha} + \sigma^2_e/K}$ with equal cluster-period size $K$.

\break
\section{Appendix: Complete scaled weight graphs of misspecified estimators}
\label{sect:appendix_weights}
Scaled weights of misspecified estimators are graphed across numbers of periods and values of $\gamma$.

\begin{figure}[htp]
    \centering
    \includegraphics[width=15cm]{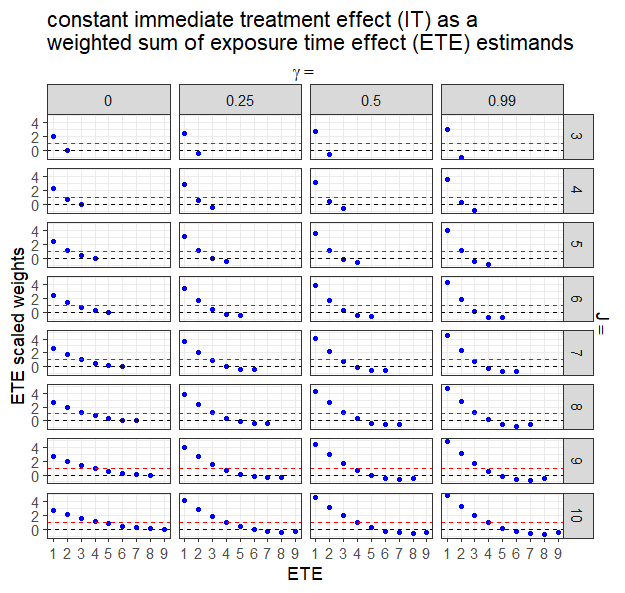}
    \caption*{Exposure time-varying treatment effects and their corresponding scaled weights $(J-1)[w_{1}(Q,\gamma,s)]$ in the IT estimator, graphed on the x-axis and y-axis, respectively. Results are presented across SW-CRTs with different total numbers of periods $J$ and varying $\gamma=\frac{\tau^{2}_{\alpha}}{\tau^{2}_{\alpha} + \sigma^2_e/K}$. Dotted red line marks a nominal weight of 1. Dotted black line marks a weight of 0.}
\end{figure}

\break
\begin{figure}[htp]
    \centering
    \includegraphics[width=15cm]{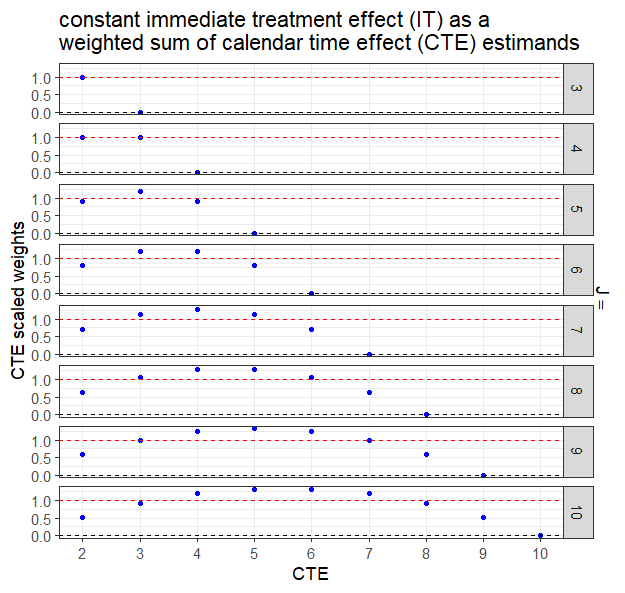}
    \caption*{Calendar time-varying treatment effects and their corresponding scaled weights $(J-2)[w_{2}(Q,j)]$ in the IT estimator, graphed on the x-axis and y-axis, respectively. Results are presented across SW-CRTs with different total numbers of periods $J$. Dotted red line marks a nominal weight of 1. Dotted black line marks a weight of 0.}
\end{figure}

\break
\begin{figure}[htp]
    \centering
    \includegraphics[width=15cm]{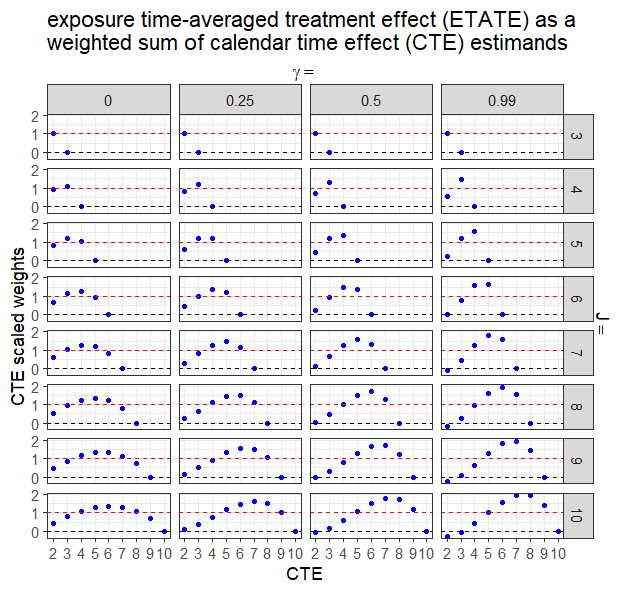}
    \caption*{Calendar time-varying treatment effects and their corresponding scaled weights $(J-2)[w_3(Q, \gamma, j)]$ in the ETATE estimator, graphed on the x-axis and y-axis, respectively. Results are presented across SW-CRTs with different total numbers of periods $J$ and varying $\gamma=\frac{\tau^{2}_{\alpha}}{\tau^{2}_{\alpha} + \sigma^2_e/K}$. Dotted red line marks a nominal weight of 1. Dotted black line marks a weight of 0.}
\end{figure}

\break
\begin{figure}[htp]
    \centering
    \includegraphics[width=15cm]{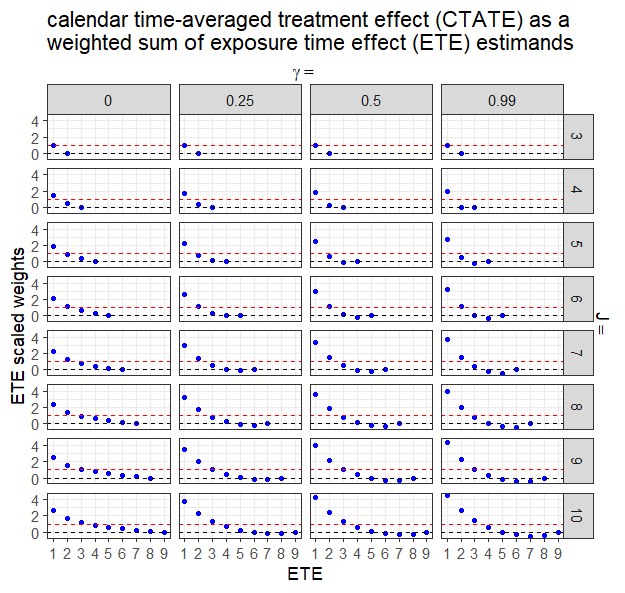}
    \caption*{Exposure time-varying treatment effects and their corresponding scaled weights $(J-2)w_4(Q,\gamma,s)$ in the CTATE estimator, graphed on the x-axis and y-axis, respectively. Results are presented across SW-CRTs with different total numbers of periods $J$ and varying $\gamma=\frac{\tau^{2}_{\alpha}}{\tau^{2}_{\alpha} + \sigma^2_e/K}$. Dotted red line marks a nominal weight of 1. Dotted black line marks a weight of 0.}
\end{figure}

\break
\section{Appendix: Additional Scenarios with CTE as analyzed by an IT estimator}
\label{sect:appendix_additional_scenarios_IT_CTE}
\begin{figure}[htp]
    \centering
    \includegraphics[width=12cm]{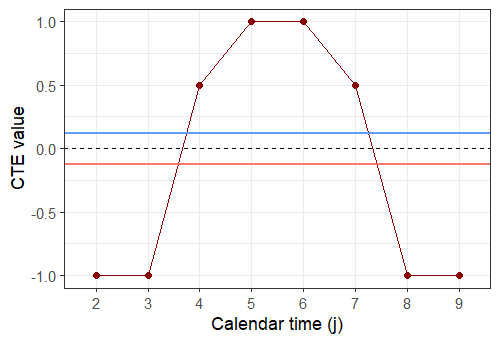}
    \caption*{An additional scenarios with different calendar time-varying treatment effects (CTE) as analyzed by an immediate treatment effect (IT) estimator in a 9 cluster, 10 period SW-CRT. CTE are only identifiable up to time period 9 with the inclusion of period fixed effects. The corresponding IT and true CTATE effect estimates are shown in the black and red lines, respectively. Dotted black line marks an effect of 0.}
\end{figure}

\break
\section{Appendix: Additional Simulation results}
\label{sect:appendix_simulation}

\begin{figure}[htp]
    \centering
    \includegraphics[width=15cm]{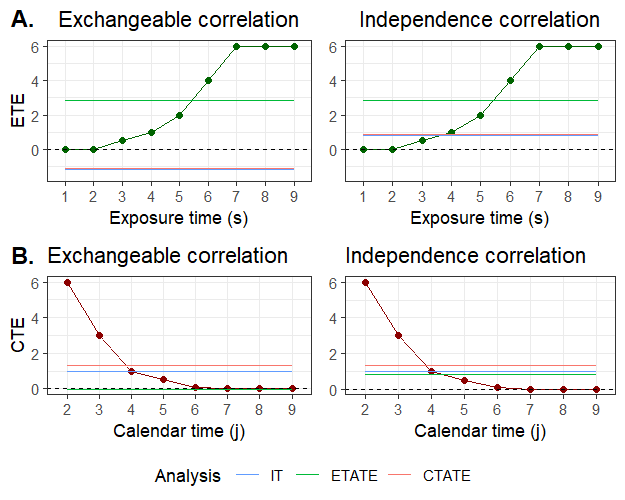}
    \caption*{Simulated (A.) exposure time-varying treatment effect (ETE) curve and (B.) calendar time-varying treatment effect (CTE) in a 18 cluster, 10 period SW-CRT with an average cluster-period cell size of 30 individuals. Results are derived with either an exchangeable or independence correlation structure. Corresponding immediate treatment effect (IT), exposure time-averaged treatment effect (ETATE), calendar time-averaged treatment effect (CTATE) estimates are plotted as horizontal lines. Analysis results are summarized over 1000 simulated replicates.}
\end{figure}

\break
\begin{figure}[htp]
    \centering
    \includegraphics[width=15cm]{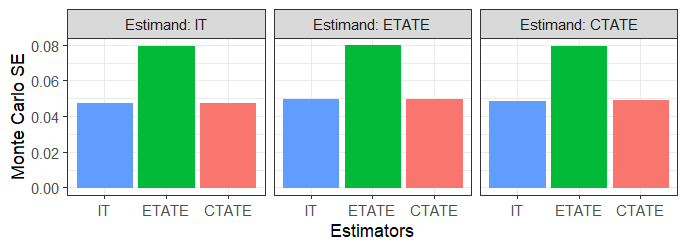}
    \caption*{Simulation results in terms of Monte Carlo SE's for calendar time-averaged treatment effect (CTATE), exposure time-averaged treatment effect (ETATE), and immediate treatment effect estimands. Corresponding results from the $\widehat{CTATE}$, $\widehat{ETATE}$, and $\widehat{IT}$ estimators are plotted on the x-axis. Analysis results are summarized over 1000 simulated replicates.}
\end{figure}

\break
\section{Appendix: Additional Case Study Re-Analysis results}
\label{sect:appendix_case_study}
\begin{figure}[htp]
    \centering
    \includegraphics[width=15cm]{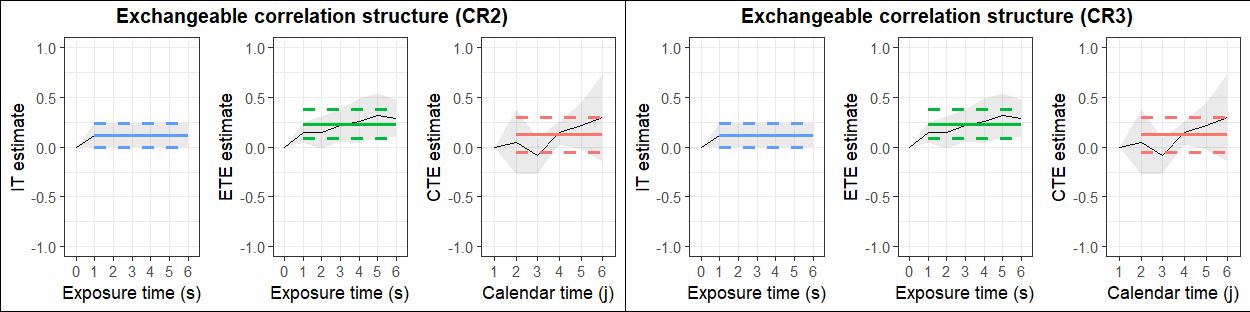}
    \caption*{Effect curves over calendar time and exposure time from an Australia disinvestment trial as analyzed with an immediate treatment, exposure time indicators, or calendar time indicators with an exchangeable correlation structure. 95\% confidence intervals of each point on the effect curve are shown in gray using bias-reduced linearization (CR2) and ``approximate jackknife'' (CR3) cluster robust variance estimators for analyses with the independence correlation structure. The black, green, and red solid lines denote the corresponding IT, ETATE, and CTATE estimates, accordingly, with dotted lines denoting the 95\% confidence intervals of the IT, ETATE, and CTATE estimates.}
\end{figure}

\end{document}